%% file: livingrev.tex
\documentclass[epsf]{livrev}
\usepackage{bm}
\usepackage{amssymb}

\begin{document} 
\title{Analytic black hole perturbation approach to gravitational radiation}

\author{Misao Sasaki and Hideyuki Tagoshi\\
{\it\small Department of Earth and Space Science, Osaka University}, \\
{\it\small Toyonaka, Osaka 560-0043, Japan}}

\date{\today}
\maketitle

\bigskip
\centerline{\large\bf Abstract}
\medskip

We review analytic methods to perform the post-Newtonian expansion of
gravitational waves induced by a particle orbiting a massive compact
body, based on the black hole perturbation theory. 
There exist two different methods of the post-Newtonian
expansion. Both are based the Teukolsky equation. 
In one method, the Teukolsky equation is transformed
into a Regge-Wheeler type equation that reduces to the 
standard Klein-Gordon equation in the flat space limit,
while in the other method, which were introduced
by Mano, Suzuki and Takasugi relatively recently, the Teukolsky 
equation is directly used in its original form. The former has 
an advantage that it is intuitively easy to understand how various
curved space effects come into play. However, it becomes increasingly
complicated when one goes on to higher and higher post-Newtonian
orders. In contrast, the latter has an advantage that a systematic
calculation to higher post-Newtonian orders is relatively
easily implementable, but otherwise so mathematical that it is hard
to understand the interplay of higher order terms.
In this paper, we review both methods so that their pros and cons
may be clearly seen. We also review some results of
calculations of gravitational radiation emitted by a particle 
orbiting a black hole.

\input livrev1.texin 
\input livrev2.texin 
\input livrev3.texin 
\input livrev4.texin 
\input livrev5.texin 
\input livrev6.texin 
\input reference.texin 
\end{document}

%% file: livrev1.texin
\section{Introduction}
\label{sec:intro}

\subsection{General}

In the past several years, there has been substantial 
progress in the projects of ground-based laser interferometric
gravitational wave detectors
which include LIGO~\cite{ref:ligoweb}, VIRGO~\cite{ref:virgoweb}, 
GEO600~\cite{ref:geoweb}, and 
TAMA300~\cite{ref:tamafirst,ref:tama2,ref:tamaweb}. 
TAMA300 has been in operation since 1999. LIGO and GEO600
started to operate in 2002. VIRGO is expected to start
operation in 2003. As another type of detectors,
resonant bar detectors~\cite{ref:bar} have been operating continuously 
for the past several years. 
There are several future projects as well. Most importantly,
the Laser Interferometer Space Antenna (LISA) project is
in progress~\cite{ref:uslisaweb,ref:eurolisaweb}.
For a review of ground and space laser interferometers, 
see, e.g., \cite{ref:detectorlivrev}.

The detection of gravitational waves will be done by extracting 
gravitational wave signals from noisy data stream. 
In developing the data analysis strategy, 
the detailed knowledge of the gravitational wave forms 
will help us greatly to detect a signal, and to extract the physical 
information of its source. 
Thus, it has become a very important problem for theorists 
to predict with sufficiently good accuracy the wave forms 
from possible gravitational wave sources. 

Gravitational waves are generated by dynamical astrophysical events
and they are expected to be strong enough to be detected when
compact stars such as neutron stars or black holes are involved in
such events. In particular, coalescing compact binaries are
considered to be the most promising sources of gravitational
radiation that can be detected by the ground-based laser 
interferometers.
The last inspiral phase of a coalescing compact binary, 
in which the binary stars orbit each other for $\sim10^4$ cycles,
will be in the bandwidth of the interferometers, and it may be
not only detected but also provides us important astrophysical
information of the system by comparing the inspiral waveform
with theoretical templates if they have sufficient accuracy.
Such an event may be detected by the initial 
phase interferometers if we are lucky, and it will be detected by the
various advanced detectors~\cite{ref:NPS,ref:phinney}. 

To predict the wave forms, a conventional approach is to 
formulate the Einstein equations with respect to the flat
Minkowski background and apply the post-Newtonian expansion
to the resulting equations (see the subsection below).

In this paper, however,
we review a different approach, namely the
black hole perturbation approach. In this approach, 
binaries are assumed to consist of a massive black hole
and a small compact star which is taken to be a point 
particle. Hence, its applicability is constrained to the case 
of binaries with large mass ratio. Nevertheless,
there are several advantages that cannot be overlooked.

Most importantly, the black hole
perturbation equations take full account of general relativistic
effects of the background spacetime and they are applicable
to arbitrary orbits of a small mass star.
In particular, if a numerical approach is taken, 
gravitational waves from highly relativistic orbits can be calculated.
Then, if we can develop a method to calculate gravitational
waves to a sufficiently high PN order analytically, 
it can not only give insight into how and when 
general relativistic effects become important, by comparing with 
numerical results, but also give us a knowledge complementary
to the conventional post-Newtonian approach, about such as 
yet-unknown higher PN order terms or general relativistic
spin effects.

Moreover, one of the main targets of LISA is to observe
phenomena associated with formation and evolution of
super-massive black holes in galactic centers. 
In particular, a gravitational wave event of a compact star
spiraling into such a super-massive black hole
is indeed the case for the black hole perturbation
theory.

\subsection{Post-Newtonian expansion of gravitational waves} 

The post-Newtonian expansion of general relativity assumes that 
the internal gravity of a source is small so that the 
deviation from the Minkowski metric is small, and that 
velocities associated with the source are small compared to the 
speed of light $c$. 
When we consider the orbital motion of a compact binary system, 
these two conditions become essentially equivalent to each other.
Although, both conditions may be violated inside each of the
compact objects, this is not regarded as a serious problem
of the post-Newtonian expansion as long as we are concerned
with gravitational waves 
generated from the orbital motion, and the two bodies are usually
assumed to be point-like objects in the calculation.
In fact, recently Itoh, Futamase, and Asada~\cite{ref:IFA1,ref:IFA2} 
developed a new
post-Newtonian method which can deal with a binary system in which
the constituent bodies may have strong internal gravity,
based on earlier work by Futamase and Schutz~\cite{ref:FuSch1,ref:FuSch2}.
They derived the equations of motion to 2.5PN order and
obtained a complete agreement with the Damour-Deruelle equations of
motion~\cite{ref:DaDe1,ref:DaDe2}, which assumes the validity of 
the point-particle approximation.

There are two existing approaches of the post-Newtonian
expansion to calculate gravitational waves: one developed by Blanchet,
Damour, and Iyer~\cite{ref:BDI,ref:blanchet} and another by
Will and Wiseman~\cite{ref:WW} based on previous work 
by Epstein, Wagoner, and Will~\cite{ref:EW,ref:WaWi}.
In both approaches, the gravitational waveforms and luminosity 
are expanded in time derivatives of radiative multipoles,
which are then related to some source multipoles (the relation
between them contains the ``tails") which are expressed as integrals
over the matter source and the gravitational field.  The source
multipoles are combined
with the equations of motion to obtain explicit expressions
in terms of the source masses, positions, and velocities. 

One issue of the post-Newtonian calculation arises from
the fact that the post-Newtonian expansion can be applied only 
to the near zone field of the source. 
In the conventional post-Newtonian formalism, the 
harmonic coordinates are used to write down the Einstein equations.
If we define the deviation from the Minkowski metric as 
\begin{equation}
h^{\mu\nu} \equiv \sqrt{-g} g^{\mu\nu}- \eta^{\mu\nu}\,,
\end{equation}
the Einstein equations are schematically written in the form 
\begin{eqnarray}
\Box h^{\mu\nu} = {16\pi} |g| T^{\mu\nu} +
\Lambda^{\mu\nu} (h), 
\label{eq:PNeinsteineq}
\end{eqnarray} 
together with the harmonic-gauge condition, $\partial_\nu h^{\mu\nu}=0$,
where $\Box =\eta^{\mu\nu} \partial_\mu\partial_\nu$, 
is the D'Alambertian operator in flat space time, 
$\eta^{\mu\nu}$ $= \rm{diag} (-1,1,1,1)$, and 
$\Lambda^{\mu\nu}(h)$ represents the non-linear terms 
in the Einstein equations. 
The Einstein equations (\ref{eq:PNeinsteineq}) are integrated using 
the flat space retarded integrals. 
In order to perform the post-Newtonian expansion, 
if we naively expand the retarded integrals in power of $1/c$, there 
appears divergent integrals.
This is a technical problem which arises due to the near zone 
nature of the post-Newtonian approximation. 
In the BDI approach, in order to 
integrate the retarded integrals, and to evaluate the 
radiative multipole moments at infinity, 
two kind of approximation method are introduced. 
One is the multipolar post-Minkowski expansion, which 
can be applied to a region outside the source including 
infinity, and the other is the near zone, post-Newtonian expansion. 
These two expansions are matched in the intermediate region where 
both expansions are valid, and the radiative multipole moments 
are evaluated at infinity. 
In the WW approach, the retarded integrals are evaluated 
directly, without expanding in terms of $1/c$, 
in the region outside the source in a novel way.

The lowest order of the gravitational waves 
are given by the Newtonian quadrupole formula. 
It is standard to call the post-Newtonian formulas for the wave forms
and luminosity which contain terms up to 
$O((v/c)^{n})$ beyond the Newtonian quadrupole formula 
as the $n/2$PN formulas.
Evaluation of gravitational waves emitted to infinity from
a compact binary system has been successfully carried
out to the 3.5 post-Newtonian (PN) order beyond the lowest
Newtonian quadrupole formula (apart from an undetermined
coefficient that appears at 3PN order) in the BDI
approach~\cite{ref:BDI,ref:blanchet}. The computation of the 3.5PN flux
requires the 3.5PN equations of motion. See a review by
Blanchet~\cite{ref:blanreview} for details on 
post-Newtonian approaches.
Up to now, both approaches give the same answer for the 
gravitational wave forms and luminosity to 2PN order.

\subsection{Linear perturbation theory of black hole}

In the black hole perturbation approach, 
we deal with gravitational waves from a particle of mass $\mu$ 
orbiting a black hole of mass $M$, assuming $\mu \ll M$. 
The perturbation of a black hole spacetime is evaluated 
to linear order in $\mu/M$. The equations are essentially
in the form of Eq.~(\ref{eq:PNeinsteineq}) with $\eta_{\mu\nu}$
replaced by the background black hole metric $g^{BH}_{\mu\nu}$
and the higher order terms $\Lambda(h)_{\mu\nu}$ neglected.
Thus, apart from the assumption $\mu\ll M$, 
the black hole perturbation approach is not restricted 
to slow-motion sources, nor to small derivations from the 
Minkowski spacetime, and the Green function used to 
integrate the Einstein equations contains 
the whole curved spacetime effect of the background geometry.

The black hole perturbation theory was originally
developed as a metric perturbation theory.
For non-rotating (Schwarzschild) black holes, a single master
equation for the metric perturbation was
derived by Regge-Wheeler~\cite{ref:RW} for the so-called
odd parity part, and later by Zerilli~\cite{ref:Zerilli}
for the even parity part. These equations have a nice property
that they reduce to the standard Klein-Gordon wave equation
in the flat space limit. However, no such equation has been
found in the case of a Kerr black hole so far.

Then, based on the Newmann-Penrose null-tetrad
formalism, in which the tetrad components of the curvature tensor 
are the fundamental variables,
a master equation for the curvature perturbation
was first developed by Bardeen and Press~\cite{ref:BaPr} 
for a Schwarzschild black hole without source 
($T^{\mu\nu}=0$), and by Teukolsky~\cite{ref:Teu} for
a Kerr black hole with source ($T^{\mu\nu}\neq 0$).
The master equation is called the Teukolsky equation, and
it is a wave equation for a null-tetrad component of the Weyl tensor
$\psi_0$ or $\psi_4$. In the source-free case,
Chrzanowski~\cite{ref:chrzanowski} and Wald~\cite{ref:Wald} 
developed a method to construct the 
metric perturbation from the curvature perturbation.

The Teukolsky equation has, however, a rather complicated structure
as a wave equation. Even in the flat space limit, it does not
reduce to the standard Klein-Gordon form. Then, Chandrasekhar
showed that the Teukolsky equation can be transformed to
the form of the Regge-Wheeler or Zerilli equation for the
source-free Schwarzschild case~\cite{ref:chandratrans}.
A generalization of this to the Kerr case with source
was done by Sasaki and Nakamura~\cite{ref:SN, ref:SN2}. They
gave a transformation that brings the Teukolsky equation
to a Regge-Wheeler type equation that reduces to the
Regge-Wheeler equation in the Schwarzschild limit.
It may be noted that the Sasaki-Nakamura equation contains
an imaginary part, suggesting that either it is unrelated to a
 (yet-to-be-found) master equation for the metric perturbation
for the Kerr geometry or it implies the non-existence of such
a master equation.

As mentioned above, an important difference between the black hole
perturbation approach and the conventional post-Newtonian approach
appears in the structure of the Green function
used to integrate the wave equations.
In the black hole perturbation, the Green function 
takes account of the curved spacetime effect on the
wave propagation, which implies complexity of its
structure in contrast to the flat space Green function.
Thus, since the system is linear in the black hole perturbation
approach, the most non-trivial task is the construction of the
Green function.

There are many papers that deal with a numerical evaluation of
the Green function and calculations of gravitational waves
induced by a particle.
See Breuer~\cite{ref:breuer}, Chandrasekhar~\cite{ref:chandra},
and Nakamura, Oohara and Kojima~\cite{ref:NOK} for reviews and
for references on earlier papers.

Here, we are interested in an analytical evaluation of
the Green function. One way is to adopt the post-Minkowski
expansion assuming $GM/c^2\ll r$.
Note that, for bound orbits, the condition $GM/c^2\ll r$ is 
equivalent to the the condition for the post-Newtonian
 expansion $v^2/c^2\ll1$.
If we can calculate the Green function to a sufficiently high order
in this expansion, we may be able to obtain a rather accurate
approximation of it that can be applicable to a 
relativistic orbit fairly close to the horizon, 
possibly to the radius as small as
the inner-most stable circular orbit (ISCO), which is given by
 $r_{ISCO}=6GM/c^2$ in the case of a Schwarzschild black hole.

It turns out that this is indeed possible. Though there arise
some complications as one goes to higher PN orders,
they are relatively easy to handle as compared to situations
one encounters in the conventional post-Newtonian approaches.
Thus, very interesting relativistic effects such as 
tails of gravitational waves can be investigated easily. 
Further, we can also easily investigate convergence properties
of the post-Newtonian expansion by comparing a numerically calculated
exact result with the corresponding analytic but approximate
result. In this sense, the analytic black hole perturbation approach
can provide an important test of the post-Newtonian expansion.

\subsection{Brief historical notes}
Let us briefly review some of past works on post-Newtonian
calculations in the black hole perturbation theory.
Although the literature on numerical
calculations of gravitational waves emitted by a particle
orbiting a black hole is abundant, there are not so many papers
that deal with the post-Newtonian expansion of gravitational waves,
mainly because it had not been necessary until recently
when the construction of accurate theoretical templates
for the interferometric gravitational wave detectors
had become an urgent issue.

In the case of orbits in the Schwarzschild background, 
one of the earliest works was done by
Gal'tsov, Matiukhin and Petukhov~\cite{ref:GMP}, who
considered a case when a particle is in a slightly eccentric orbit 
around a Schwarzschild black hole, and calculated the gravitational 
waves up to 1PN order. 
Poisson~\cite{ref:poisson} considered a circular orbit 
around a Schwarzschild black hole and calculated 
the wave forms and luminosity to 1.5PN order 
at which the tail effect appears. 
Cutler, Finn, Poisson and Sussman~\cite{ref:CFPS} also worked on
the same problem numerically by using the least square fitting 
technique to the numerically evaluated data for the luminosity, 
and obtained a post-Newtonian formula 
for the luminosity to 2.5PN order. 
Subsequently, a highly accurate numerical calculation was 
carried out by Tagoshi and Nakamura~\cite{ref:TN}. They obtained the 
formulas for the luminosity to 4PN order numerically by using 
the least square fitting method. They found the 
$\log v$ terms in the luminosity formula at 3PN and 4PN orders. 
They concluded that, although the convergence 
of the post-Newtonian expansion is slow, the luminosity formula
accurate to 3.5PN order will be good enough as theoretical
templates for the ground-based interferometers,
to represent the orbital phase evolution of 
coalescing compact binaries.
After that, Sasaki~\cite{ref:sasaki} found an analytic method 
and obtained formulas which are needed to 
calculate the gravitational waves
to 4PN order. Then, 
Tagoshi and Sasaki~\cite{ref:TS} obtained the 
gravitational wave forms and 
luminosity to 4PN order analytically, and confirmed the results of 
Tagoshi and Nakamura. 
These calculations were extended to 5.5PN order by Tanaka, Tagoshi, 
and Sasaki~\cite{ref:TTS}.

In the case of orbits around a Kerr black hole, 
Poisson calculated the 1.5PN order corrections to the wave forms and
luminosity due to the rotation of the black hole and showed that the
result agrees with the standard post-Newtonian effect due to
spin-orbit coupling~\cite{ref:poisson2}.
Then, Shibata, Sasaki, Tagoshi and Tanaka~\cite{ref:SSTT} 
calculated the luminosity to 2.5PN order. They 
calculated the luminosity from a particle in circular orbit 
with small inclination from the equatorial plane. 
They used the Sasaki-Nakamura equation as well as the 
Teukolsky equation. 
This analysis was extended to 4PN order by 
Tagoshi, Shibata, Tanaka and Sasaki~\cite{ref:TSTS}
in which the orbit of the test particle was restricted to
circular ones on the equatorial plane. 
The analysis in the case of slightly eccentric orbit on the 
equatorial plane was also done by Tagoshi~\cite{ref:tagoshi} 
to 2.5PN order. 

Tanaka, Mino, Sasaki and Shibata~\cite{ref:TMSS} 
considered the case when a spinning particle 
is in a circular orbit near the equatorial plane 
of a Kerr black hole, based on the Papaetrou equations of motion
for a spinning particle~\cite{ref:papa}
and the energy momentum tensor of a spinning particle
 by Dixon~\cite{ref:dixon}.
They derived the luminosity formula to 2.5PN
order which includes the linear order effect of the particle's spin.

The absorption of gravitational waves into the black hole horizon, 
appearing at 4PN order in the Schwarzschild case, 
was calculated by Poisson and Sasaki for a particle 
in a circular orbit~\cite{ref:PoSa}. 
The black hole absorption in the case of rotating black hole
appears at 2.5PN order~\cite{ref:galtsov}. 
Using a new analytic method to solve the homogeneous Teukolsky
equation found by Mano, Suzuki and Takasugi~\cite{ref:MST},
the black hole absorption in the Kerr case
was calculated by Tagoshi, Mano, and Takasugi~\cite{ref:TMT}
to 6.5PN order beyond the quadrupole formula. 

If gravity is not described by the Einstein theory but by the
Brans-Dicke theory, there will appear scalar 
type gravitational waves as well as transverse-traceless 
gravitational waves. Such scalar type gravitational waves were
calculated to 2.5PN order by Ohashi, Tagoshi and Sasaki~\cite{ref:OTS}
in the case when a compact star is in a circular orbit on the
equatorial plane around a Kerr black hole.

In the rest of the paper, we use the units $c=G=1$. 

%% file: livrev2.texin
\section{Basic formulas for the black hole perturbation}
\label{sec:teukolsky}

\subsection{Teukolsky formalism}
In terms of the conventional Boyer-Lindquist coordinates, the metric of
a Kerr black hole is expressed as
\begin{eqnarray}
\label{Kerr}
ds^2&=&-{\Delta\over\Sigma}(dt-a\sin^2\theta\,d\varphi)^2
+{\sin^2\theta\over\Sigma}\left[(r^2+a^2)d\varphi-a\,dt\right]^2
\nonumber\\
&&+{\Sigma\over\Delta}dr^2+\Sigma d\theta^2\,,
\end{eqnarray}
where $\Sigma=$ $r^2+a^2\cos^2\theta$ and $\Delta=$ $r^2-2Mr+a^2$.
In the Teukolsky formalism~\cite{ref:Teu}, 
the gravitational perturbations of a Kerr black 
hole are described by a Newman-Penrose quantity 
$\psi_4=-C_{\alpha\beta\gamma\delta}
n^{\alpha}\bar{m}^\beta n^\gamma \bar{m}^\delta$, where
$C_{\alpha\beta\gamma\delta}$
is the Weyl tensor and 
\begin{eqnarray*}
&&
n^\alpha=
((r^2+a^2),-\Delta,0,a)/(2\Sigma),\nonumber\\
&&
m^\alpha=
(ia\sin\theta,0,1,i/\sin\theta)/(\sqrt{2}(r+ia\cos\theta)).
\end{eqnarray*}
The perturbation equation for $\phi\equiv \rho^{-4}\psi_4$, 
$\rho=(r-ia\cos\theta)^{-1}$, is given by 
\begin{equation}
{}_s{\cal O}\,\phi=4\pi\Sigma \hat{T}\,. \label{eq:masterTeu1}
\end{equation}
Here, the operator ${}_s{\cal O}$ is given by
\begin{eqnarray}
{}_s{\cal O}
&=&-\left[{(r^2+a^2)^2\over \Delta}-a^2 \sin^2\theta\right]
\partial_t^2-{4Mar\over \Delta}\partial_t\partial_\phi
-\left[{a^2\over \Delta}-{1\over \sin^2\theta}\right]
\partial_\phi^2 \nonumber\\
&&+\Delta^{-s}\partial_r(\Delta^{s+1}\partial_r)+
{1\over \sin\theta}\partial_\theta(\sin\theta\partial_\theta)+
2s\left[{a(r-M)\over \Delta}+{i\cos\theta\over \sin^2\theta}
\right]\partial_\phi \nonumber\\
&&
+2s\left[{M(r^2-a^2)\over \Delta}-r-ia\cos\theta\right]\partial_t
-s(s\cot^2\theta-1), \label{eq:masterTeu2}
\end{eqnarray}
with $s=-2$. 
The source term $\hat T$ is given by
\begin{eqnarray}
\hat{T}&=&2(B'_2+B_2^{*'})\,,
\nonumber
\\
B_2'&=&-{1 \over 2}\rho^8{\overline \rho}L_{-1}[\rho^{-4}L_0
(\rho^{-2}{\overline \rho}^{-1}T_{nn})]  \nonumber\\
&-&{1 \over 2\sqrt{2}}\rho^8{\overline \rho}\Delta^2 L_{-1}[\rho^{-4}
{\overline \rho}^2 J_+(\rho^{-2}{\overline \rho}^{-2}\Delta^{-1}
T_{{\overline m}n})],  \nonumber\\
B_2'^*&=&-{1 \over 4}\rho^8{\overline \rho}\Delta^2 J_+[\rho^{-4}J_+
(\rho^{-2}{\overline \rho}T_{{\overline m}{\overline m}})] \nonumber\\
&-&{1 \over 2\sqrt{2}}\rho^8{\overline \rho}\Delta^2 J_+[\rho^{-4}
{\overline \rho}^2 \Delta^{-1} L_{-1}(\rho^{-2}{\overline \rho}^{-2}
T_{{\overline m}n})], \label{eq:B2}\\
\nonumber
\end{eqnarray}
where
\begin{eqnarray}
L_s&=&\partial_{\theta}+{m \over \sin\theta}
-a\omega\sin\theta+s\cot\theta, \label{eq:rho}\\
J_+&=&\partial_r+{iK/\Delta}\,;\quad
K=(r^2+a^2)\omega-ma\,,
\nonumber
\end{eqnarray}
and $T_{nn}$, $T_{{\overline m}n}$ and 
$T_{{\overline m}{\overline m}}$  
are the tetrad components of the energy momentum tensor
($T_{nn}=T_{\mu\nu}n^\mu n^\nu$ etc.).
The bar denotes the complex conjugation.

If we set $s=2$ in Eq.~(\ref{eq:masterTeu1}), with appropriate
change of the source term, it becomes 
the perturbation equation for $\psi_0$.
Moreover, it describes the perturbation for scalar field ($s=0$), 
neutrino field ($|s|=1/2$), and electromagnetic field ($|s|=1$) as well. 

We decompose $\psi_4$ into the Fourier-harmonic components according to
\begin{equation}
\rho^{-4}\psi_4=
\sum_{\ell m}\int d\omega 
e^{-i\omega t+i m \varphi} \ _{-2}S_{\ell m}(\theta)
R_{\ell m\omega}(r), \label{eq:psifour}
\end{equation}
The radial function $R_{\ell m\omega}$ and the angular function 
$_{s}S_{\ell m}(\theta)$ satisfy the Teukolsky 
equations with $s=-2$ as 
\begin{equation}
\Delta^2{d\over dr}\left({1\over \Delta}{dR_{\ell m\omega}\over dr}
\right)
-V(r) R_{\ell m\omega}=T_{\ell m\omega}, 
\label{eq:radTeukolsky}
\end{equation}
\begin{eqnarray}
& &\Bigl[{1 \over \sin\theta}{d \over d\theta}
 \Bigl\{\sin\theta {d \over d\theta} \Bigr\}
-a^2\omega^2\sin^2\theta
 -{(m-2\cos\theta)^2 \over \sin^2\theta}
\nonumber\\
& &~~~~~~~~~~~~~~~
+4a\omega\cos\theta-2+2ma\omega+\lambda\Bigr] 
{}_{-2}S_{\ell m}=0. \label{eq:spheroid}
\end{eqnarray}
The potential $V(r)$ is given by
\begin{equation}
V(r)=-{K^2+4i(r-M)K \over \Delta}+8i\omega r+\lambda,
\end{equation} 
where $\lambda$ is the eigenvalue 
of $_{-2}S_{\ell m}^{a\omega}$.
The angular function $_{s}S_{\ell m}(\theta)$ is called 
the spin-weighted spheroidal harmonic, which is usually normalized as 
\begin{equation}
\int_0^{\pi} |_{-2}S_{\ell m}|^2 \sin\theta d\theta=1.
\label{eq:snorm}
\end{equation}
In the Schwarzschild limit, it reduces to the spin-weighted spherical
harmonic with $\lambda\to\ell(\ell+1)$. In the Kerr case, however,
no analytic formula for $\lambda$ is known.
The source term $T_{\ell m\omega}$ is given by 
\begin{equation}
T_{\ell m\omega}
 =4\int d\Omega dt\rho^{-5}{\overline\rho}^{-1}(B_2'+B_2'^*)
e^{-im\varphi+i\omega t}{_{-2}S^{a\omega}_{\ell m} \over
\sqrt{2\pi}},
\label{eq:source}
\end{equation}
We mention that for orbits of our interest, which are bounded,
$T_{\ell m\omega}$ has support only in a compact range of $r$.

We solve the radial Teukolsky equation by using the 
Green function method. 
For this purpose, 
we define two kinds of homogeneous solutions of the radial 
Teukolsky equation: 
\begin{eqnarray}
R^{\rm in}_{\ell m\omega}
&\to&
\cases{
B^{\rm trans}_{\ell m\omega}\Delta^2  e^{-ikr^*}\,&
for $r\to r_+$ \cr
r^3 B^{\rm  ref}_{\ell m\omega}e^{i\omega r^*}+
r^{-1}B^{\rm inc}_{\ell m\omega} e^{-i\omega r^*}\,&
for $r\to +\infty,$ \cr}
\label{eq:Rin} 
\\
~\nonumber\\
R^{\rm up}_{\ell m\omega}
&\to&
\cases{ 
C^{\rm  up}_{\ell m\omega} e^{ik r^*}+ 
\Delta^2 C^{\rm ref}_{\ell m\omega} e^{-ik r^*}\,& 
for $r\to r_+,$ \cr
C^{\rm trans}_{\ell m\omega} r^3  e^{i\omega r^*}\,& 
for $r\to +\infty$\cr 
} 
\label{eq:Rup} 
\end{eqnarray} 
where $k=\omega-ma/2Mr_+$ and 
$r^*$ is the tortoise coordinate defined by 
\begin{eqnarray}
r^{*}&=&\int {dr^*\over dr}dr \nonumber\\
&=&r+{2Mr_{+}\over {r_{+}-r_{-}}}\ln{{r-r_{+}}\over 2M}
-{2Mr_{-}\over {r_{+}-r_{-}}}\ln{{r-r_{-}}\over 2M},
\label{eq:rst}
\end{eqnarray}
where $r_{\pm}=M\pm \sqrt{M^2-a^2}$, and we
 have fixed the integration constant. 

Combining with the Fourier mode $e^{-i\omega t}$, we see that
$R^{\rm in}_{\ell m\omega}$ has no outcoming wave from past horizon,
while $R^{\rm up}$ has no incoming wave at past infinity.
Since these are the property of waves causally generated by
a source, a solution of the Teukolsky equation 
which has purely outgoing property at
infinity and has purely ingoing property at the horizon 
is given by 
\begin{equation}
R_{\ell m\omega}={1\over W_{\ell m\omega}}
 \left\{R^{\rm up}_{\ell m\omega}\int^r_{r_+}dr' 
 R^{\rm in}_{\ell m\omega}
T_{\ell m\omega}\Delta^{-2} 
+ R^{\rm in}_{\ell m\omega}\int^\infty_{r}dr' 
R^{\rm up}_{\ell m\omega} T_{\ell m\omega}\Delta^{-2}\right\}, 
\end{equation}
where the Wronskian $W_{\ell m\omega}$ is given by 
\begin{equation}
W_{\ell m\omega}=2 i \omega C^{\rm trans}_{\ell m\omega}
     B^{\rm inc}_{\ell m\omega}. 
\end{equation}
Then, the asymptotic behavior at the horizon is 
\begin{equation}
R_{\ell m\omega}(r\to r_+)\to 
{{B^{\rm trans}_{\ell m\omega} \Delta^2 e^{-i k r^*}} \over 
2i\omega C^{\rm trans}_{\ell m\omega}B^{\rm inc}_{\ell m\omega}}
\int^{\infty}_{r_+}dr' R^{\rm up}_{\ell m\omega} 
    T_{\ell m\omega}\Delta^{-2}
\equiv \tilde{Z}^{\rm H}_{\ell m\omega} \Delta^2 e^{-i k r^*}\,, 
\label{eq:zhole}
\end{equation}
while the asymptotic behavior at infinity is
\begin{equation}
R_{\ell m\omega}(r\to\infty)
 \to{r^3e^{i\omega r^*} \over 2i\omega 
   B^{\rm inc}_{\ell m\omega}}
\int^{\infty}_{r_+}dr'{T_{\ell m\omega}(r') 
R^{\rm in}_{\ell m\omega}(r') 
\over\Delta^{2}(r')}
\equiv \tilde Z_{\ell m\omega}^{\infty}r^3e^{i\omega r^*}\,.
\label{eq:Rinfty}
\end{equation}

We note that the homogeneous Teukolsky equation is invariant 
under the complex conjugation followed by the transformation
$m\rightarrow -m$ and $\omega\rightarrow -\omega$. 
Thus, we can set $\bar{R}^{\rm in,up}_{\ell m\omega}$ 
$=R^{\rm in,up}_{\ell -m -\omega}$,
where the bar denote the complex conjugation. 

We consider $T_{\mu\nu}$ of a monopole particle of mass $\mu$.
The energy momentum tensor takes the form,
\begin{equation}
T^{\mu\nu}
={\mu\over \Sigma \sin\theta dt/d\tau}{dz^{\mu}\over d\tau}
{dz^{\nu}\over d\tau}\delta(r-r(t))\delta(\theta-\theta(t))
  \delta(\varphi-\varphi(t)).
\end{equation}
where $z^\mu=\bigl(t,r(t),\theta(t),\varphi(t)\bigr)$ is a geodesic
trajectory and $\tau=\tau(t)$ is the proper time along the geodesic.
The geodesic equations in the Kerr geometry are given by
\begin{eqnarray}
& &\Sigma{d\theta \over d\tau}
      =\pm\Bigl[C-\cos^2\theta \Bigl\{a^2(1-E^2)+
      {l_z^2 \over \sin^2\theta}\Bigr\}\Bigr]^{1/2} 
      \equiv \Theta(\theta),\nonumber\\
& &\Sigma{d\varphi \over d\tau}
      =-\Bigl(aE-{l_z \over \sin^2\theta}\Bigr)
      +{a \over \Delta}\Bigl(E(r^2+a^2)-al_z\Bigr) 
       \equiv \Phi, \nonumber\\
& &\Sigma{dt \over d\tau}=
      -\Bigl(aE-{l_z \over \sin^2\theta}\Bigr)a\sin^2\theta
      +{r^2+a^2 \over \Delta}\Bigl(E(r^2+a^2)-al_z\Bigr) 
      \equiv T, \nonumber\\
& &\Sigma{dr \over d\tau}=\pm\sqrt{R},  
\label{eq:geod}\\
\nonumber
\end{eqnarray}
where $E$, $l_z$ and $C$ are the energy, the $z$-component of the
angular momentum and the Carter constant of a test particle,
respectively,\footnote{These constants of 
motion are those measured in units of $\mu$. 
That is, if expressed in the standard units, $E$, $l_z$ and $C$ in 
Eq.~(\ref{eq:geod}) are to be replaced 
with $E/\mu$, $l_z/\mu$ and $C/\mu^2$,
respectively.}
and
\begin{equation}
R=[E(r^2+a^2)-al_z]^2-\Delta[(Ea-l_z)^2+r^2+C].
\label{eq:Rgeo}
\end{equation}

Using Eq.~(\ref{eq:geod}), the tetrad components of the energy momentum 
tensor are expressed as
\begin{eqnarray}
T_{n\,n}&=&\mu{C_{n\,n} \over \sin\theta}
\delta(r-r(t)) \delta(\theta-\theta(t)) \delta(\varphi-\varphi(t)),
\nonumber\\
T_{{\overline m}\,n}&=&\mu{C_{{\overline m}\,n} \over \sin\theta}
\delta(r-r(t)) \delta(\theta-\theta(t)) \delta(\varphi-\varphi(t)),
\label{eq:tnn}\\
T_{{\overline m}\,{\overline m}}&=&\mu
{C_{{\overline m}\,{\overline m}} \over \sin\theta}
\delta(r-r(t)) \delta(\theta-\theta(t)) \delta(\varphi-
\varphi(t)),\nonumber\\
\nonumber
\end{eqnarray}
where
\begin{eqnarray}
C_{n\,n}&=&{1\over 4\Sigma^3 \dot t}\left[E(r^2+a^2)-al_z
+\Sigma{dr\over d\tau} \right]^2,\nonumber\\
C_{{\overline m}\,n}&=&
-{\rho \over 2\sqrt{2}\Sigma^2 \dot t}\left[E(r^2+a^2)-al_z
+\Sigma{dr\over d\tau} \right]
\left[i\sin\theta\Bigl(aE-{l_z \over \sin^2\theta}\Bigr)
+\Sigma \frac{d\theta}{d\tau}\right], 
\nonumber\\
C_{{\overline m}\,{\overline m}}&=&
{\rho^2 \over 2\Sigma \dot t }
\left[i\sin\theta 
\Bigl(aE-{l_z \over \sin^2\theta}\Bigr)
+\Sigma \frac{d\theta}{d\tau}\right]^2,
\label{eq:cnn}
\end{eqnarray}
and $\dot t=dt/d\tau$.
Substituting Eq.~(\ref{eq:B2}) into Eq.~(\ref{eq:source}) and 
performing integration by part, we obtain
\begin{eqnarray}
T_{\ell m\omega}
&=&
{4\mu\over\sqrt{2\pi}}\int^{\infty}_{-\infty}
dt\int d\theta e^{i\omega t-im\varphi(t)} \nonumber\\
&&
\times\Bigl[-{1 \over 2}L_1^{\dag} \bigl\{ \rho^{-4}L_2^{\dag}(\rho^3 S)
\bigr\}
C_{n\,n}\rho^{-2}{\overline \rho}^{-1}\delta(r-r(t))
\delta(\theta-\theta(t)) \nonumber\\
&&
+{\Delta^2 {\overline \rho}^2 \over \sqrt{2} \rho}
\bigl(L_2^{\dag} S+ia({\overline \rho}-\rho)\sin\theta S\bigr)
J_+ \bigg\{ 
{C_{{\overline m}\,n}\over \rho^{2}{\overline \rho}^{2}\Delta}
\delta(r-r(t))\delta(\theta-\theta(t)) \bigg\} \nonumber\\
&&
+{1 \over 2\sqrt{2} }
L_2^{\dag}\bigl\{ \rho^3 S({\overline \rho}^2 \rho^{-4})_{,r} \bigr\}
C_{{\overline m}\,n}\Delta \rho^{-2}{\overline \rho}^{-2}
\delta(r-r(t))\delta(\theta-\theta(t)) \nonumber\\
&&
-{1 \over 4}\rho^3 \Delta^2 S J_+\bigl\{\rho^{-4}
J_+\bigl({\overline \rho} \rho^{-2}C_{{\overline m}\,{\overline m}}
\delta(r-r(t))\delta(\theta-\theta(t))\bigr) \bigr\}
\Bigr], 
\label{eq:tass2}
\end{eqnarray}
where
\begin{equation}
L_s^{\dag}=\partial_{\theta}-{m \over \sin\theta}
+a\omega\sin\theta+s\cot\theta, 
\end{equation}
and $S$ denotes $_{-2}S_{\ell m}^{a\omega}(\theta)$ for simplicity.

For a source bounded in a finite range of $r$,
it is convenient to rewrite Eq.~(\ref{eq:tass2}) further as
\begin{eqnarray}
T_{\ell m\omega}&=&\mu\int^{\infty}_{-\infty}dt 
e^{i\omega t-i m \varphi(t)}
\Delta^2\Bigl[(A_{n\,n\,0}+A_{{\overline m}\,n\,0}+
A_{{\overline m}\,{\overline m}\,0})\delta(r-r(t)) \nonumber\\
&+&\left\{(A_{{\overline m}\,n\,1}+A_{{\overline m}\,{\overline m}\,1})
\delta(r-r(t))\right\}_{,r}
+\left\{A_{{\overline m}\,{\overline m}\,2}
\delta(r-r(t))\right\}_{,rr}\Bigr],
\label{eq:tone2} \\
\nonumber
\end{eqnarray}
where 
\begin{eqnarray}
A_{n\,n\,0}&=&{-2 \over \sqrt{2\pi}\Delta^2}
C_{n\,n}\rho^{-2}{\overline \rho}^{-1}
L_1^+\{\rho^{-4}L_2^+(\rho^3 S)\},\nonumber \\
A_{{\overline m}\,n\,0}&=&{2 \over \sqrt{\pi}\Delta} 
C_{{\overline m}\,n}\rho^{-3}
\Bigl[\left(L_2^+S\right)
\Bigl({iK \over \Delta}+\rho+{\overline \rho}\Bigr)
\nonumber \\
& &\qquad-a\sin\theta S {K \over \Delta}
   ({\overline \rho}-\rho)\Bigr],\nonumber \\
A_{{\overline m}\,{\overline m}\,0}
&=&-{1 \over \sqrt{2\pi}}\rho^{-3}{\overline \rho}
C_{{\overline m}\,{\overline m}}S\Bigl[
-i\Bigl({K \over \Delta}\Bigr)_{,r}-{K^2 \over \Delta^2}+
2i\rho {K \over \Delta}\Bigr],\nonumber \\
A_{{\overline m}\,n\,1}&=&{2\over \sqrt{\pi}\Delta }
\rho^{-3}C_{{\overline m}\,n}
[L_2^+S+ia\sin\theta({\overline \rho}-\rho)S],\nonumber \\
A_{{\overline m}\,{\overline m}\,1}
&=&-{2 \over \sqrt{2\pi}}
\rho^{-3}{\overline \rho}
C_{{\overline m}\,{\overline m}}S
\Bigl(i{K \over \Delta}+\rho\Bigr),\nonumber \\
A_{{\overline m}\,{\overline m}\,2}
&=&-{1\over \sqrt{2\pi}}\rho^{-3}{\overline \rho}
C_{{\overline m}\,{\overline m}}S. 
\label{eq:ann} 
\end{eqnarray}
Inserting Eq.~(\ref{eq:tone2}) into Eq.~(\ref{eq:Rinfty}), we obtain 
$\tilde Z_{\ell m\omega}$ as
\begin{equation}
\tilde Z_{\ell m\omega}=
{\mu\over2i\omega B^{\rm inc}_{\ell m\omega}}
\int^{\infty}_{-\infty}dt e^{i\omega t-i m \varphi(t)}
W_{\ell m\omega}\,,
\label{eq:zzq2}
\end{equation}
where 
\begin{eqnarray}
W_{\ell m\omega}&=&
\Bigl[R^{\rm in}_{\ell m\omega}\{A_{n\,n\,0}+A_{{\overline m}\,n\,0}
+A_{{\overline m}\,{\overline m}\,0}\}
\nonumber\\
& &-{dR^{\rm in}_{\ell m\omega} \over dr}\{ A_{{\overline m}\,n\,1}
+A_{{\overline m}\,{\overline m}\,1}\}
 +{d^2 R^{\rm in}_{\ell m\omega} \over dr^2}
 A_{{\overline m}\,{\overline m}\,2}\Bigr]_{r=r(t)}\,.
\label{eq:Wdef}
\end{eqnarray}

In this paper, we focus on orbits which are either circular (with or
without inclination) or eccentric but confined on the equatorial
plane. In either case, the frequency spectrum of $T_{\ell m\omega}$
becomes discrete. Accordingly, $\tilde Z_{\ell m\omega}$ in
Eq.~(\ref{eq:zhole}) or (\ref{eq:Rinfty}) takes the form,
\begin{equation}
  \label{eq:tildeZ}
  \tilde Z_{\ell m\omega}=\sum_n \delta(\omega-\omega_n)
    Z_{\ell m\omega}\,.
\end{equation}
Then, in particular, $\psi_4$ at $r \to \infty$ 
is obtained from Eq.~(\ref{eq:psifour}) as
\begin{equation}
\psi_4={1\over r}\sum_{\ell m n}
Z_{\ell m\omega_n}{{}_{-2}S_{\ell m}^{a\omega_n} \over \sqrt{2\pi}}
e^{i\omega_n(r^*-t)+im\varphi}. 
\label{eq:psi4inf} 
\end{equation}
At infinity, $\psi_4$ is related to the two independent modes of
gravitational waves $h_+$ and $h_\times$ as
\begin{equation}
\psi_4={1\over2}(\ddot h_{+}-i\ddot h_{\times}). 
\label{eq:hplcr} 
\end{equation}
{}From Eqs.~(\ref{eq:psi4inf}) and (\ref{eq:hplcr}), 
the luminosity averaged over $t \gg \Delta t$, where $\Delta t$ is the
characteristic time scale of the orbital motion (e.g., a period between
the two consecutive apastrons), is given by 
\begin{equation}
\left\langle{dE \over dt}\right\rangle=\sum_{\ell,m,n} 
{\Bigl|Z_{\ell m\omega_n}\Bigr|^2\over4\pi\omega_n^2} 
\equiv \sum_{\ell,m,n}\Bigl( {dE \over dt} \Bigr)_{\ell mn}.
\label{eq:Eflux} 
\end{equation}
In the same way, the time-averaged angular momentum flux is given by
\begin{equation}
\left\langle{dJ_z \over dt}\right\rangle=\sum_{\ell,m,n} 
{m\Bigl| Z_{\ell m\omega_n}\Bigr|^2\over4\pi\omega_n^3}
\equiv \sum_{\ell,m,n}\Bigl( {dJ_z \over dt} \Bigr)_{\ell mn}
=\sum_{\ell,m,n}{m\over\omega_n}\Bigl( {dE \over dt} \Bigr)_{\ell mn}\,.
\label{eq:Jflux} 
\end{equation}

\subsection{Chandrasekhar-Sasaki-Nakamura transformation}

As seen from the asymptotic behaviors of the radial functions
given in Eqs.~(\ref{eq:zhole}) and (\ref{eq:Rinfty}),
the Teukolsky equation is not in the form of a canonical wave equation
near the horizon and infinity. Therefore, it is desirable to find
a transformation that brings the radial Teukolsky equation into 
the form of a standard wave equation.

In the Schwarzschild case, Chandrasekhar found that
the Teukolsky equation can be transformed to the Regge-Wheeler
equation, which has the standard form of a wave equation with solutions
having regular asymptotic behaviors at horizon and 
infinity~\cite{ref:chandratrans}.
 The Regge-Wheeler equation was originally derived as an
equation governing the odd parity metric perturbation~\cite{ref:RW}. 
The existence of this transformation implies that the Regge-Wheeler equation
can describe the even parity metric perturbation simultaneously,
though the explicit relation of the Regge-Wheeler function obtained
by the Chandrasekhar transformation with the actual metric perturbation
variables has not been given in the literature yet. 

Later, Sasaki and Nakamura succeeded in generalizing the Chandrasekhar 
transformation to the Kerr case~\cite{ref:SN, ref:SN2}.
The Chandrasekhar-Sasaki-Nakamura
transformation was originally introduced
to make the potential in the radial equation short-ranged and 
to make the source term well-behaved at horizon and infinity.
Since we are interested only in bound orbits, it is not necessary 
to perform this transformation. Nevertheless, because its flat space limit
reduces to the standard radial wave equation in the Minkowski spacetime,
it is convenient to apply the transformation when dealing with
the post-Minkowski or post-Newtonian expansion, at least at low
orders of expansion.

We transform the homogeneous Teukolsky equation to the
Sasaki-Nakamura equation~\cite{ref:SN, ref:SN2}, which is given by 
\begin{equation}
\Bigl[{d^2 \over dr^{*2}}-F(r){d \over dr^{*}}-U(r)\Bigr]
X_{\ell m\omega}=0. 
\label{eq:sneq}
\end{equation}
The function $F(r)$ is given by
\begin{equation}
F(r)={\eta_{,r} \over \eta}{\Delta \over r^2+a^2},
\label{eq:Fdef}
\end{equation}
where 
\begin{equation}
\eta=c_0+c_1/r+c_2/r^2+c_3/r^3+c_4/r^4,
\label{eq:etadef}
\end{equation}
with
\begin{eqnarray}
&c_0&=-12i\omega M+\lambda(\lambda+2)-12a \omega(a\omega-m),\nonumber\\
&c_1&=8ia[3a\omega-\lambda(a\omega-m)],\nonumber\\
&c_2&=-24ia M(a\omega-m)+12a^2[1-2(a\omega-m)^2 ],\nonumber\\
&c_3&=24ia^3(a\omega-m)-24Ma^2,\nonumber\\
&c_4&=12a^4.\label{eq:ceq}
\end{eqnarray}
The function $U(r)$ is given by
\begin{equation}
U(r)={\Delta U_1\over(r^2+a^2)^2}+G^2+{\Delta G_{,r}\over r^2+a^2}-FG,
\label{eq:Udef}
\end{equation}
where
\begin{eqnarray}
G&=&-{2(r-M) \over r^2+a^2}+{r\Delta \over (r^2+a^2)^2},\nonumber\\
U_1&=&V+{\Delta^2 \over \beta}
\Bigl[\Bigl(2\alpha+{\beta_{,r}  \over \Delta}\Bigr)_{,r}
-{\eta_{,r} \over \eta}
\Bigl(\alpha +{\beta_{,r} \over \Delta}\Bigr)\Bigr],\nonumber\\
\alpha&=&-i{K \beta \over \Delta^2}+3iK_{,r}
+\lambda+{6\Delta \over r^2},\nonumber\\
\beta&=&2\Delta\Bigl(-iK+r-M-{2\Delta \over r}\Bigr).\label{eq:ur}
\end{eqnarray}

The relation between $R_{\ell m\omega}$
and $X_{\ell m\omega}$ is 
\begin{equation}
R_{\ell m\omega}={1 \over \eta}
\Bigl\{\Bigl(\alpha+{\beta_{,r} \over \Delta}\Bigr)\chi_{\ell m\omega}
-{\beta \over \Delta}\chi_{\ell m\omega,r} \Bigr\},
\label{eq:xtor}
\end{equation}
where $\chi_{\ell m\omega}=X_{\ell m\omega} \Delta/(r^2+a^2)^{1/2}$.
Conversely, we can express $X_{\ell m\omega}$ in terms of 
$R_{\ell m\omega}$ as
\begin{equation}
X_{\ell m\omega}=(r^2+a^2)^{1/2}\,r^2\,J_{-}J_{-}
\left[{1\over r^2}R_{\ell m\omega}\right], 
\label{eq:rtox}
\end{equation}
where $J_{-}=(d/dr)-i(K/\Delta)$. 

If we set $a=0$, this transformation reduces to the 
Chandrasekhar transformation for the 
Schwarzschild black hole~\cite{ref:chandratrans}. 
The explicit form of the transformation is
\begin{eqnarray}
R_{\ell m\omega}&=&{\Delta\over c_0}\Big({d\over dr^*}+i\omega\Big)
{r^2\over \Delta}\Big({d\over dr^*}+i\omega\Big)(r X_{\ell m\omega}), 
\\
X_{\ell m\omega}&=&{r^5\over \Delta}
\Big({d\over dr^*}-i\omega\Big)
{r^2\over \Delta}\Big({d\over dr^*}+i\omega\Big)
{R_{\ell m\omega}\over r^2}, 
\end{eqnarray}
where $c_0$, defined in Eq.~(\ref{eq:ceq}), reduces to
$c_0=(\ell-1)\ell(\ell+1)(\ell+2)-12iM\omega$. 
In this case, the Sasaki-Nakamura equation (\ref{eq:sneq}) 
reduces to the Regge-Wheeler equation~\cite{ref:RW} 
which is given by 
\begin{equation}
\left[{{d^2\over dr^{*2}}+\omega^2-V(r)}\right]X_{\ell\omega}(r)
=0, \label{eq:rw}
\end{equation}
where 
\begin{equation}
V(r)=\left({1-{2M\over r}}\right)\left({{\ell(\ell+1)\over r^2}
-{6M\over r^3} }\right). 
\end{equation}
As clear from the above form of the equation, 
the lowest order solutions are given by the spherical Bessel functions.
Hence it is intuitively straightforward to apply
the post-Newtonian expansion to it. Some useful techniques for the
post-Newtonian expansion were developed for the Schwarzschild case
by Poisson~\cite{ref:poisson} and Sasaki~\cite{ref:sasaki}.

The asymptotic behavior of 
the ingoing-wave solution $X^{{\rm in}}$ which corresponds
to Eq.~(\ref{eq:Rin}) is 
\begin{equation}
X^{\rm in}_{\ell m\omega}\to\cases{
A^{\rm ref}_{\ell m\omega} e^{i\omega r^*}+
A^{\rm inc}_{\ell m\omega} e^{-i\omega r^*}\,& 
for $r^* \to \infty$ \cr
A^{\rm trans}_{\ell m\omega} e^{-ik r^*}\, & for
$r^* \to -\infty$. }
\label{eq:xinasy}
\end{equation}
The coefficients $A^{\rm inc}$, $A^{\rm ref}$ and 
$A^{\rm trans}$ are related to 
$B^{\rm inc}$, $B^{\rm ref}$ and 
$B^{\rm trans}$, defined in Eq.~(\ref{eq:Rin}), by 
\begin{eqnarray}
B^{\rm inc}_{\ell m\omega} &=&
   -{1\over 4\omega^2}A^{\rm inc}_{\ell m\omega}, 
\label{eq:ainbin}\\
B^{\rm ref}_{\ell m\omega}&=&-{4\omega^2\over c_0}
A^{\rm ref}_{\ell m\omega}, 
\label{eq:arefbref}\\
B^{\rm trans}_{\ell m\omega}&=&
 {1\over d_{\ell m\omega}}A^{\rm trans}_{\ell m\omega}, 
\label{eq:atransbtrans}
\end{eqnarray}
where
\begin{eqnarray*}
d_{\ell m\omega}&=&\sqrt{2Mr_+}[(8-24iM\omega-16M^2\omega^2)r_{+}^2 \\
&&+(12iam-16M+16amM\omega+24iM^2\omega)r_{+}\\
&&-4a^2m^2-12iamM+8M^2]. 
\end{eqnarray*}

In the following sections, we present a method
of post-Newtonian expansion based on the above formalism in
the case of the Schwarzschild background.
In the Kerr case, although a post-Newtonian expansion method developed
in previous work~\cite{ref:SSTT,ref:TSTS}
was based on the Sasaki-Nakamura equation,
we will not present it in this paper.
Instead, we present a different formalism, namely the one developed by
Mano, Suzuki and Takasugi which allows us to solve 
the Teukolsky equation in a more systematic manner,
albeit very mathematical~\cite{ref:MST}.
The reason is that the equations in the Kerr case
are already complicated enough even if one uses the Sasaki-Nakamura
equation, so that there is not much advantage in using it.
In contrast, in the Schwarzschild case,
it is much easier to obtain physical insight into the role of
relativistic corrections if we deal with the Regge-Wheeler equation.

%% file: livrev3.texin
\section{Post-Newtonian expansion of the Regge-Wheeler equation}
\label{sec:ReggeWheeler}

In this section, we review a post-Newtonian expansion method for
the Schwarzschild background, based on the Regge-Wheeler equation.
We focus on the gravitational waves emitted to infinity, but
not on those absorbed by the black hole. The black hole absorption
is deferred to Section~\ref{sec:MSTformalism}, in which we review
the Mano-Suzuki-Takasugi method for solving the Teukolsky equation.

Since we are interested in the waves emitted to infinity,
as seen from Eq.~(\ref{eq:Rinfty}), what we need is a method 
to evaluate the ingoing-wave Teukolsky
function $R^{\rm in}_{\ell m\omega}$, or its counterpart
of the Regge-Wheeler equation, $X^{\rm in}_{\ell m\omega}$,
which are related by Eq.~(\ref{eq:xtor}).
In addition, we assume $\omega>0$ whenever it is necessary
throughout this section. Formulas and equations for $\omega<0$
are obtained from the symmetry 
$\bar{X}^{\rm in}_{\ell m\omega}=X^{\rm in}_{\ell -m-\omega}$.

\subsection{Basic assumptions}

We consider the case of a test particle with mass $\mu$ in an orbit
which is nearly circular around a black hole with mass 
$M\gg\mu$. 
For a nearly circular orbit, say at $r\sim r_0$,
what we need to know is the behavior of
 $R^{\rm in}_{\ell m\omega}$ at $r\sim r_0$. 
In addition, the contribution of $\omega$ to $R^{\rm in}_{\ell m\omega}$ 
comes mainly from $\omega\sim m\Omega_\varphi$, where
$\Omega_\varphi\sim (M/r_0^3)^{1/2}$ is the orbital angular frequency.

Thus, if we express the Regge-Wheeler
equation (\ref{eq:rw}) in terms of a non-dimensional variable
$z\equiv \omega r$, with a non-dimensional parameter
$\epsilon\equiv 2M\omega$, we are interested in the behavior
of $X^{\rm in}_{\ell m\omega}(z)$ at
$z\sim\omega r_0\sim m (M/r_0)^{1/2}\sim v$ with
 $\epsilon\sim 2m (M/r_0)^{3/2}$ $\sim v^3$, where
$v\equiv (M/r_0)^{1/2}$ is the characteristic orbital velocity.
The post-Newtonian expansion assumes that $v$ is much smaller
than the velocity of light; $v\ll 1$. Consequently, we have
$\epsilon\ll v\ll 1$ under the post-Newtonian expansion.

To obtain $X^{\rm in}_{\ell m\omega}$ (which we denote below by
$X_\ell$ for simplicity) under these assumptions,
we find it convenient to rewrite the Regge-Wheeler equation
in an alternative form. It is
\begin{equation}
\Big[{d^2\over dz^2}+{2\over z}{d\over dz}
+\Big(1-{\ell(\ell+1)\over z^2}\Big)\Big]\xi_\ell
=\epsilon e^{-iz}{d\over dz}\left[{1\over z^3}
   {d\over dz}\left(e^{iz}z^2\xi_\ell(z)\right)\right],
\label{eq:xieq}
\end{equation}
where $\xi_\ell$ is a function related to $X_\ell$ as
\begin{equation}
X_{\ell}=ze^{-i\epsilon\ln(z-\epsilon)}\xi_{\ell}\,.
\label{eq:xidef}
\end{equation}
The ingoing-wave boundary condition of $\xi_\ell$ is derived
 from Eqs.~(\ref{eq:xinasy}) and (\ref{eq:xidef}) as 
\begin{equation}
\xi_{\ell}\to\cases{
A^{\rm inc}_\ell e^{i\epsilon\ln\epsilon} z^{-1} e^{-iz}
+ A^{\rm ref}_\ell e^{-i\epsilon\ln\epsilon} z^{-1}
e^{i(z+ 2\epsilon \ln z)}\,& 
for $r^* \to \infty$, \cr
A^{\rm trans}_\ell \epsilon^{-1}
e^{i \epsilon (\ln \epsilon - 1)}\, & for
$r^* \to -\infty$. }
\label{eq:xiasy}
\end{equation}
The above form of the Regge-Wheeler equation is used
in the following subsections.

It should be noted that if we recover the gravitational constant
$G$, we have $\epsilon=2GM\omega$. Thus, the expansion in terms of 
$\epsilon$ corresponds to the post-Minkowski expansion, and
expanding the Regge-Wheeler equation
with the assumption $\epsilon\ll 1$ gives a set of iterative
wave equations on the flat spacetime background. 
One of the most significant differences between the 
black hole perturbation theory and any theory based on 
the flat spacetime background is the presence of 
the black hole horizon in the former case. 
Thus, if we naively expand the Regge-Wheeler equation 
with respect to $\epsilon$, the horizon boundary condition
becomes unclear, since there is no horizon on the
flat spacetime. To establish the boundary condition at 
the horizon, we need to treat the Regge-Wheeler equation 
near the horizon separately. We thus have to find a
solution near the horizon and the solution
obtained by the post-Minkowski expansion must be
matched with it in the region where both solutions
are valid.

It may be of interest to note the difference between 
the matching used in the BDI approach for the
post-Newtonian expansion~\cite{ref:blanchet,ref:BDI}
and the matching used here. 
In the BDI approach, the matching is done between 
the post-Minkowskian metric and the near-zone post-Newtonian metric.
In our case, the matching is done between the post-Minkowskian
gravitational field and the gravitational field near
the black hole horizon. 

\subsection{Horizon solution; $\bm{z\ll 1}$}

In this subsection, we first consider the solution
near the horizon, which we call the horizon solution,
based on \cite{ref:PoSa}.
To do so, we assume
$z \ll 1$ and treat $\epsilon$ as a small 
number, but leave the ratio $z/\epsilon$ arbitrary.
We change the independent variable to $x=1-z/\epsilon$
and the wave function to
\begin{eqnarray}
Z =\left({\epsilon\over z}\right)^3{X_\ell\over A^{\rm trans}_\ell}
=\left({\epsilon\over z}\right)^2 {\epsilon\,\xi_\ell\over
A^{\rm trans}_\ell e^{i \epsilon (\ln \epsilon - 1)}}\,.
\end{eqnarray}
Note that the horizon corresponds to $x=0$.
We then have
\begin{eqnarray}
&&x(x-1)Z'' + \bigl[2(3-i\epsilon)x-(1-2i\epsilon)\bigr] Z'
\nonumber \\ 
&&\qquad
+\bigl[6-\ell(\ell+1)-5i\epsilon+
\epsilon^2(2-3x+x^2) \bigr] Z = 0,
\label{eq:Zeq}
\end{eqnarray}
where a prime denotes differentiation with respect to $x$. 
We look for a solution which is regular at $x=0$.

First, we consider the lowest order solution
by setting $\epsilon=0$ in Eq.~(\ref{eq:Zeq}).
The boundary condition (\ref{eq:xiasy}) 
requires that $Z=1$ at $x=0$.
The solution which satisfies the boundary condition is 
\begin{eqnarray}
Z=\sum_{n=0}^{\ell-2} {(2-\ell)_n (\ell+3)_n\over n!} x^n\,;
\quad
(a)_n={\Gamma(a+n)\over \Gamma(a)}. 
\label{eq:Zeq0}
\end{eqnarray}
Thus, the lowest order solution is a polynomial of order $\ell-2$
in $x=1-z/\epsilon$. 

Next, we consider the solution accurate to $O(\epsilon)$. 
We neglect the terms of $O(\epsilon^2)$ in Eq.~(\ref{eq:Zeq}).
Then its takes the form of a hypergeometric equation, 
\begin{equation}
x(x-1)Z'' + \bigl[(a+b+1)x-c\bigr] Z'+abZ=0, 
\end{equation}
with parameters
\begin{eqnarray}
a &=& -(\ell-2) - i \epsilon + O(\epsilon^2), \nonumber \\
b &=& \ell+3-i \epsilon + O(\epsilon^2), \label{eq:hypara} \\
c &=& 1-2i\epsilon. \nonumber
\end{eqnarray}
The two linearly independent solutions are $F(a,b;c;x)$ and
$x^{1-c} F(a+1-c,b+1-c;2-c;x)$, where $F$ is the hypergeometric
function. However, only the first solution
is regular at $x=0$. 
We therefore obtain
\begin{equation}
\xi_\ell(z) = A^{\rm trans}_\ell\epsilon^{-1}
e^{i \epsilon (\ln\epsilon - 1)} \left({z\over\epsilon}\right)^2
F\left(a,b;c;1-{z\over\epsilon}\right).
\label{eq:xihorizon}
\end{equation}

The above solution must be matched with the solution obtained
from the post-Minkowski expansion of Eq.~(\ref{eq:xieq}),
which we call the outer solution, in a region both solutions are valid.
It is the region where the post-Newtonian expansion is applied,
i.e., the region $\epsilon\ll z\ll1$.
For this purpose, we rewrite Eq.~(\ref{eq:xihorizon}) 
as (see, e.g., Eq.~(15.3.8) of \cite{ref:handbook})
\begin{eqnarray}
\xi_\ell&=&
A^{\rm trans}_\ell\epsilon^{-1}
e^{i\epsilon(\ln\epsilon-1)}
\biggl[\left({z\over \epsilon}\right)^{\ell+i\epsilon}
\frac{\Gamma(c) \Gamma(b-a)}{\Gamma(b)\Gamma(c-a)}
F(a,c-b;a-b+1;{\epsilon\over z})
\cr
&&\qquad
+\left({z\over \epsilon}\right)^{-\ell-1+i\epsilon}
\frac{\Gamma(c) \Gamma(a-b)}{\Gamma(a)\Gamma(c-b)}
F(b,c-a;b-a+1;{\epsilon\over z})\biggr]. 
\label{eq:xihorizon2}
\end{eqnarray}
This naturally allows the expansion in $\epsilon/z$.
It should be noted that 
the second term in the square brackets of the above expression
is meaningless as it is, since the factor $\Gamma(a-b)$
diverges for integer $\ell$.
So, when evaluating the second term,
we first have to extend $\ell$ to a non-integer number. Then,
only after expanding it in terms of $\epsilon$, we should take 
the limit of an integer $\ell$. One then finds that this 
procedure gives rise to an additional factor of $O(\epsilon)$.
For $\epsilon/z\ll1$, it therefore becomes
$O(\epsilon^{2\ell+2})$ higher in $\epsilon$ than the first term.
Then we obtain
\begin{eqnarray}
\xi_\ell(\epsilon \ll z \ll 1) &=&
\frac{(2\ell)!}{(\ell-2)!(\ell+2)!}
\frac{z^\ell}{\epsilon^{\ell+1}}
\biggl[1 + i \epsilon (a_\ell +\ln z)
\nonumber \\ 
& & \qquad
-\frac{(\ell-2)(\ell+2)}{2\ell} \frac{\epsilon}{z} +
O(\epsilon^2) \biggr],
\label{eq:xihorizon3}
\end{eqnarray}
where 
\begin{equation}
a_\ell = 2\gamma + \psi(\ell-1) + \psi(\ell+3) - 1, 
\end{equation}
and $\psi(n)$ is the digamma function,
\begin{equation}
\psi(n)=-\gamma+\sum_{k=1}^{n-1} k^{-1},
\end{equation}
and $\gamma \simeq 0.57721$ is the Euler constant.

As we will see below, the above solution is accurate enough
to determine the boundary condition
of the outer solution up to 6PN order of expansion.

\subsection{Outer solution; $\bm{\epsilon \ll 1}$}

We now solve Eq.~(\ref{eq:xieq}) in the limit $\epsilon \ll 1$,
i.e., by applying the post-Minkowski expansion to it.
In this subsection, we consider the solution to $O(\epsilon)$.
Then  we match the solution to the horizon solution given by
Eq.~(\ref{eq:xihorizon3}) at $\epsilon\ll z\ll1$.

By setting 
\begin{equation}
\xi_\ell(z) = \sum_{n=0}^{\infty} \epsilon^n \xi_\ell^{(n)}(z), 
\label{eq:xiexpand}
\end{equation}
each $\xi_\ell^{(n)}(z)$ is found to satisfy
\begin{equation}
\Big[{d^2\over dz^2}+{2\over z}{d\over dz}
+\Big(1-{\ell(\ell+1)\over z^2}\Big)\Big]\xi_\ell^{(n)}
=e^{-iz}{d\over dz}\left[{1\over z^3}
   {d\over dz}\left(e^{iz}z^2\xi_\ell^{(n-1)}(z)\right)\right].
\label{eq:xieq2}
\end{equation}
Equation (\ref{eq:xieq2}) is an inhomogeneous spherical Bessel
equation. It is the simplicity of this equation which motivated
the introduction of the auxiliary function $\xi_\ell$~\cite{ref:sasaki}.

The zeroth-order solution, $\xi_\ell^{(0)}$, satisfies the homogeneous 
spherical Bessel equation, and must be a linear combination 
of the spherical Bessel functions of the first and 
second kinds, $j_\ell(z)$ and $n_\ell(z)$. 
Here, we demand the compatibility with the horizon solution
(\ref{eq:xihorizon3}). Since $j_\ell(z)\sim z^\ell$ and 
$n_\ell(z)\sim z^{-\ell-1}$, $n_\ell(z)$ does not match 
with the horizon solution at the leading order of $\epsilon$. 
We therefore have 
\begin{equation}
\xi_\ell^{(0)}(z) = \alpha_\ell^{(0)} j_\ell(z).
\label{eq:xi0}
\end{equation}
The constant $\alpha_\ell^{(0)}$ represents the overall 
normalization of the solution. Since it can be chosen
arbitrary, we set $\alpha_\ell^{(0)}=1$ below.

The procedure to
obtain $\xi_\ell^{(1)}(z)$ was described in detail in
\cite{ref:sasaki}. 
Using the Green function $G(z,z') = j_\ell(z_<) n_\ell(z_>)$, 
Eq.~(\ref{eq:xieq2}) may be put into an indefinite integral
form,
\begin{eqnarray}
\xi_\ell^{(n)}&=&n_\ell\int^z dz z^2 e^{-iz}j_\ell
\left[{1\over z^3}(e^{iz}z^2 \xi_\ell^{(n-1)}(z))'\right]'\cr
&&
-j_\ell\int^z dz z^2 e^{-iz}n_\ell
\left[{1\over z^3}(e^{iz}z^2 \xi_\ell^{(n-1)}(z))'\right]'.
\label{eq:xigreen}
\end{eqnarray}
The calculation is tedious but straightforward.
All the necessary formulas to obtain
$\xi^{(n)}_{\ell}$ for $n\le2$ are given in Appendix 
of \cite{ref:sasaki} or Appendix D of \cite{ref:supplement}.
Using those formulas, we have for $n=1$,
\begin{eqnarray}
\xi^{(1)}_{\ell}
&=&
\alpha_\ell^{(1)}j_\ell+\beta_\ell^{(1)} n_\ell \cr
&&
+{(\ell-1)(\ell+3)\over2(\ell+1)(2\ell+1)}j_{\ell+1}
    -\left({\ell^2-4\over2\ell(2\ell+1)}
        +{2\ell-1\over\ell(\ell-1)}\right)j_{\ell-1}
\cr
&&
+R_{\ell,0} j_0
+\sum_{m=1}^{\ell-2}\left({1\over m}+{1\over m+1}\right)
           R_{\ell,m} j_m
-2D_{\ell}^{nj}+ij_\ell\ln z. 
\label{eq:xi1}
\end{eqnarray}
Here, $D_{\ell}^{nj}$ and $R_{\ell,m}$ are functions
defined as follows. The function $D_{\ell}^{nj}$ is given by
\begin{equation}
D_{\ell}^{nj}={1\over 2}\left[j_{\ell} \,{\rm Si}(2z)-n_{\ell} 
\left({\rm Ci}(2z)-\gamma-\ln 2z\right)\right], 
\end{equation}
where ${\rm Ci}(x)=-\int^{\infty}_x dt\cos t/t$ and 
${{\rm Si}(x)}=\int^x_0 dt\sin t/t$. 
The function $R_{m,k}$
is defined by $R_{m,k} =z^2 (n_m j_k -j_m n_k)$.
It is a polynomial in inverse powers of $z$ given by
\begin{equation}
R_{m,k}= -\sum_{r=0}^{[(m -k -1)/2]} (-1)^r
    {\Gamma(m-k-r)\Gamma\left(m+{1\over 2}-r \right)
    \over r!\,\Gamma(m -k -2r)\Gamma\left(k +{3\over 2}+r\right)}
    \left({2\over z}\right)^{m-k-1-2r},    
\label{eq:Rmkdef}
\end{equation}
for $m>k$, and 
\begin{equation}
 R_{m,k}=-R_{k,m}.
\label{eq:Rmksym}
\end{equation}
for $m<k$. 

Here, we again perform the matching with the horizon solution
(\ref{eq:xihorizon3}). It should be noted that $\xi^{(1)}_\ell$
given by Eq.~(\ref{eq:xi1}) is regular in the limit $z\to0$
except for the term $\beta^{(1)}_\ell n_\ell$.
By examining the asymptotic behavior 
of Eq.~(\ref{eq:xi1}) at $z\ll 1$, we find $\beta_\ell^{(1)}=0$,
i.e., the solution is regular at $z=0$.
As for $\alpha_\ell^{(1)}$, it only contributes 
to the renormalization of $\alpha_\ell^{(0)}$. Hence we 
set $\alpha_\ell^{(1)}=0$. 
Thus, the transmission amplitude $A^{\rm trans}_\ell$ is determined 
to $O(\epsilon)$ as 
\begin{equation}
A^{\rm trans}_\ell
={(\ell-2)!(\ell+2)!\over (2\ell)!(2\ell+1)!}\epsilon^{\ell+1}
[1-i\epsilon\,a_\ell+O(\epsilon^2)]\,.
\end{equation}
It may be noted that this explicit expression for $A^{\rm trans}_\ell$
is unnecessary for the evaluation of gravitational waves at infinity.
It is relevant only for the evaluation
of the black hole absorption.

\subsection{More on the inner boundary condition
of the outer solution}

In this subsection, we discuss the inner boundary 
condition of the outer solution in more detail. 
As we have seen in the previous subsection, 
the boundary condition of $\xi_{\ell}$ 
is that it is regular at $z\to0$,
at least to $O(\epsilon)$. While in the full
non-linear level, the horizon boundary is at $z=\epsilon$.
We therefore investigate to what order in $\epsilon$
the condition of regularity at $z=0$ can be applied.

Let us consider the general form of
the horizon solution. With $x=1-z/\epsilon$, it
is expanded in the form,
\begin{equation}
\xi_{\ell}
=\xi_{\ell}^{\{0\}}(x)+\epsilon\, \xi_{\ell}^{\{1\}}(x)
          +\epsilon^2\, \xi_{\ell}^{\{2\}}(x)+\cdots. 
\end{equation}
The lowest order solution $\xi_{\ell}^{\{0\}}(x)$ is given 
by the polynomial (\ref{eq:Zeq0}). 
Apart from the common overall factor, it is 
schematically expressed as
\begin{eqnarray}
\xi_{\ell}^{\{0\}}
=\left(z\over \epsilon\right)^\ell
\bigg[1+c_1{\epsilon\over z}+\cdots +
c_{\ell-2}\bigg({\epsilon\over z}\bigg)^{\ell-2}\bigg].
\end{eqnarray}
Thus, $\xi_{\ell}^{\{0\}}$ does not have a term matched
with $n_\ell$, but it matches with $j_\ell$. 
We have
 $\xi_{\ell}^{\{0\}}=z^\ell\epsilon^{-\ell}\sim\epsilon^{-\ell}j_\ell$. 
A term that matches with $n_\ell$ first appears
from $\xi_{\ell}^{\{1\}}$. This can be seen from 
the horizon solution valid to $O(\epsilon)$, Eq.~(\ref{eq:xihorizon2}).
The second term in the square brackets of it produces a term 
$\epsilon\,(z/\epsilon)^{-\ell-1}=\epsilon^{\ell+2}z^{-\ell-1}
\sim\epsilon^{\ell+2}n_\ell$. 
This term therefore becomes $O(\epsilon^{2\ell+2}/z^{2\ell+1})$ 
higher than the lowest order term $\epsilon^{-\ell}j_\ell$. 
Since $\ell\geq 2$, this effect first appears at $O(\epsilon^6)$
in the post-Minkowski expansion, while
it first appears at $O(v^{13})$ in the post-Newtonian expansion
if we note that $\epsilon=O(v^3)$ and $z=O(v)$.
This implies, in particular, that if we are interested in the
gravitational waves emitted to infinity, we may simply
impose the regularity at $z=0$ as the inner boundary
condition of the outer solution for the calculation
up to 6PN order beyond the quadrupole formula.

The above fact that a non-trivial boundary condition due to the
presence of the black hole horizon appears at $O(\epsilon^{2\ell+2})$
in the post-Minkowski expansion can be more easily seen
as follows. Since $j_\ell=O(z^\ell)$ as $z\rightarrow0$, we have
$X_\ell\rightarrow O(\epsilon^{\ell+1})e^{-iz^*}$,
or $A^{\rm trans}_\ell=O(\epsilon^{\ell+1})$,
where $z^*=z+\epsilon\ln(z-\epsilon)$.
On the other hand, from the asymptotic behavior of $j_\ell$ at
$z=\infty$, the coefficients $A^{\rm inc}_\ell$ and $A^{\rm ref}_\ell$ 
must be of order unity. Then using the Wronskian argument,
we find
\begin{equation}
 |A^{\rm inc}_\ell|-|A^{\rm ref}_\ell|
={|A^{\rm trans}_\ell|^2\over|A^{\rm inc}_\ell|+|A^{\rm ref}_\ell|}
    =O(\epsilon^{2\ell+2}).
\end{equation}
Thus, we immediately see that a non-trivial boundary condition
appears at $O(\epsilon^{2\ell+2})$.

It is also useful to keep in mind the above fact when we solve
for $\xi_\ell$ under the post-Minkowski expansion.
It implies that we may choose a phase so that
$A^{\rm inc}_\ell$ and $A^{\rm ref}_\ell$ are complex conjugate 
to each other, to $O(\epsilon^{2\ell+1})$.
With this choice, the imaginary part of $X_\ell$, which reflects the 
boundary condition at horizon, do not appear until
$O(\epsilon^{2\ell+2})$ because the Regge-Wheeler equation is real. 
Then recalling the relation of $\xi_\ell$ to $X_\ell$,
Eq.~(\ref{eq:xidef}), ${\rm Im}\left(\xi^{(n)}_\ell\right)$
for a given $n\le 2\ell+1$ is completely determined in terms of 
${\rm Re}\left(\xi^{(r)}_\ell\right)$ for $r\le n-1$.
That is, we may focus on solving
only the real part of Eq.~(\ref{eq:xieq2}).

\subsection{Structure of the ingoing-wave function to
 $\bm{O(\epsilon^2)}$} 

With the boundary condition discussed in the previous 
section, we can integrate the ingoing-wave
Regge-Wheeler function iteratively to higher orders of $\epsilon$ 
in the post-Minkowskian expansion, $\epsilon\ll 1$. 
This was carried out in \cite{ref:sasaki} to $O(\epsilon^2)$
and in \cite{ref:TTS} to $O(\epsilon^3)$
(See \cite{ref:supplement} for details).
Here, we do not recapitulate the details of the calculation
since it is already quite involved at $O(\epsilon^2)$,
with much less space for physical intuition.
Instead, we describe the general properties of the 
ingoing-wave function to $O(\epsilon^2)$. 

As discussed in the previous section, 
the ingoing wave Regge-Wheeler function $X_\ell$ can be 
made real up to $O(\epsilon^{2\ell+1})$, or to $O(\epsilon^5)$
of the post-Minkowski expansion if we recall $\ell\geq2$.
Choosing the phase of $X_\ell$ in this way, 
let us explicitly write down the expressions of
 ${\rm Im}(\xi^{(n)}_\ell)$ ($n=1,2$) in terms of
 ${\rm Re}(\xi_\ell^{(m)})$ ($m\leq n-1$). 
We decompose the real and imaginary parts of 
$\xi_\ell^{(n)}$ as 
\begin{equation}
\xi_\ell^{(n)}=f_\ell^{(n)}+ig_\ell^{(n)}.
\end{equation}
Inserting this expression into Eq.~(\ref{eq:xidef}) and 
expanding the result with respect to $\epsilon$, 
and noting $f_\ell^{(0)}=j_\ell$ and $g_\ell^{(0)}=0$, 
we find 
\begin{eqnarray}
X_\ell&=&
e^{-i\epsilon\ln(z-\epsilon)}z(j_\ell+
\epsilon(f_\ell^{(1)}+ig_\ell^{(1)})+
\epsilon^2(f_\ell^{(2)}+ig_\ell^{(2)})+\cdots)
\nonumber\\
&=&
z\left(j_\ell+\epsilon f_\ell^{(1)}
+\epsilon^2\left(f_\ell^{(2)}+g_\ell^{(1)}\ln z-
{1\over 2}j_\ell(\ln z)^2\right)+\cdots\right)
\nonumber\\
&&+iz\left(\epsilon(g_\ell^{(1)}-j_\ell\ln z)+
\epsilon^2\left(g_\ell^{(2)}+{1\over z}j_\ell-
f_\ell^{(1)}\ln z\right)+\cdots\right).
\end{eqnarray}
Hence, we have 
\begin{equation}
g_\ell^{(1)}=j_\ell\,\ln z\,
\quad
g_\ell^{(2)}=-{1\over z}\,j_\ell+f_\ell^{(1)}\ln z, 
\cdots.
\end{equation}
We thus have the post-Minkowski expansion of $X_\ell$ as 
\begin{eqnarray}
&&
X_\ell=\sum_{n=0}^{\infty} \epsilon^n X_\ell^{(n)},
\nonumber\\
&&
X_\ell^{(0)}=z\, j_\ell\,,
\quad
X_\ell^{(1)}=z f_\ell^{(1)}\,,
\quad
X_\ell^{(2)}=z\left(f_\ell^{(2)}
+{1\over 2}j_\ell\,(\ln z)^2\right), \cdots.
\label{eq:Xinge}
\end{eqnarray}

Now, let us consider the asymptotic behavior of 
$X_\ell$ at $z\ll 1$. As we know that $\xi^{(1)}_\ell$
and $\xi^{(2)}_\ell$ are regular at $z=0$, 
it is readily obtained by simply assuming 
Taylor expansion forms for them (including possible
$\ln z$ terms), inserting them to Eq.~(\ref{eq:xieq2}),
and comparing the terms of the same order on both
sides of the equation.
We denote the right-hand side of Eq.~(\ref{eq:xieq2}) 
by $S^{(n)}_\ell$.

For $n=1$, we have
\begin{eqnarray}
{\rm Re}\left(S^{(1)}_\ell\right)
&=&
{1\over z}\left(j_\ell''+{1\over z}j_\ell'
                       -{4+z^2\over z^2}j_\ell\right)
\nonumber\\
&=&
\cases{O(z)~&for $\ell=2$,\cr
        O(z^{\ell-3})~&for $\ell\geq 3$.\cr}
\label{eq:S1}
\end{eqnarray}
Inserting this to Eq.~(\ref{eq:xieq2}) with $n=1$
we find
\begin{equation}
{\rm Re}\left(\xi^{(1)}_\ell\right)=f^{(1)}_\ell
=\cases{O(z^3)~&for $\ell=2$,\cr
        O(z^{\ell-1})~&for $\ell\geq3$.\cr}
\label{eq:Rexione}
\end{equation}
Of course, this behavior is consistent
with the full post-Minkowski solution
given in Eq.~(\ref{eq:xi1}).

For $n=2$, we then have
\begin{eqnarray}
{\rm Re}\left(S^{(2)}_\ell\right)
&=&{1\over z}\left(f^{(1)}_\ell{}''+{1\over z}f^{(1)}_\ell{}'
                       -{4+z^2\over z^2}f^{(1)}_\ell\right)
      -{1\over z}\left(2g^{(1)}_\ell{}'+{1\over z}g^{(1)}_\ell\right)
\nonumber\\
&=&
-{1\over z}\left(j_\ell\,\ln z\right)'-{1\over z^2}j_\ell\,\ln z
+\cases{O(z^{\ell-2})~&for $\ell=2,3$,\cr
        O(z^{\ell-4})~&for $\ell\geq4$,\cr}
\label{eq:S2}
\end{eqnarray}
This gives
\begin{equation}
{\rm Re}\left(\xi^{(2)}_\ell\right)=f^{(2)}_\ell
   =\cases{O(z^\ell)+O(z^\ell)\ln z
                     -{1\over2}j_\ell(\ln z)^2~&for $\ell=2,3$,\cr
           O(z^{\ell-2})+O(z^\ell)\ln z
                     -{1\over2}j_\ell(\ln z)^2~&for $\ell\geq4$.\cr}
\label{eq:Rexitwo}
\end{equation}
Note that the $\ln z$ terms in (\ref{eq:S2}) arising from 
$g^{(1)}_\ell$ gives the $(\ln z)^2$ term in $f^{(2)}_\ell$
that just cancels the $j_\ell(\ln z)^2/2$ term of
$X^{(2)}_\ell$ in Eq.~(\ref{eq:Xinge}).

Inserting Eqs.~(\ref{eq:Rexione}) and (\ref{eq:Rexitwo}) 
into the relevant expressions in Eq.~(\ref{eq:Xinge}), we find
\begin{eqnarray}
X_2&=&
z^3\left[O(1)+\epsilon\, O(z)
          +\epsilon^2\left\{O(1)+O(1)\ln z\right\}+\cdots\right],
\cr
X_3&=&
z^3\left[O(z)+\epsilon\, O(1)
          +\epsilon^2\left\{O(z)+O(z)\ln z\right\}+\cdots\right],
\label{eq:Xinz0}
\\
X_\ell&=&
z^3\left[O(z^{\ell-2})+\epsilon\, O(z^{\ell-3})
          +\epsilon^2\left\{O(z^{\ell-4})
                   +O(z^{\ell-2})\ln z\right\}
                    +\cdots\right]~(\ell\geq4).
\nonumber
\end{eqnarray}
Note that, for $\ell=2$ and $3$,
the leading behavior of $X_\ell^{(n)}$ at $n=\ell-1$ 
is more regular than the naively expected behavior, $\sim z^{\ell+1-n}$,
which propagates to the consecutive higher order terms in $\epsilon$.
This behavior seems to hold for general $\ell$, but we do not know a
physical explanation of it.

Given a post-Newtonian order to which we want to calculate, 
by setting $z=O(v)$ and $\epsilon=O(v^3)$, the above asymptotic
behaviors tell us the highest order of $X_\ell^{(n)}$ we need.
We also see the presence of $\ln z$ terms in $X_\ell^{(2)}$. 
The logarithmic terms appear as a consequence of the 
mathematical structure of the Regge-Wheeler equation at $z\ll 1$. 
The simple power series expansion of $X_\ell^{(n)}$ in terms of $z$
breaks down at $O(\epsilon^2)$, and we have to add logarithmic
terms to obtain the solution. 
These logarithmic terms will give rise to $\ln v$ terms in the 
wave-form and luminosity formulas at infinity, beginning 
at $O(v^6)$~\cite{ref:TN,ref:TS}. 
It is not easy to explain physically how these $\ln v$ terms appear. 
But the above analysis suggests that the $\ln v$ terms 
in the luminosity originate from some spatially 
local curvature effects in the near zone. 

Now we turn to the asymptotic behavior at $z=\infty$.
For this purpose, let the asymptotic form of $f^{(n)}_\ell$ be
\begin{equation}
f^{(n)}_\ell\rightarrow P^{(n)}_\ell j_\ell+Q^{(n)}_\ell n_\ell
 \quad{\rm as}~z\rightarrow\infty\,.
\label{eq:PQdef}
\end{equation}
Noting Eq.~(\ref{eq:Xinge}) and the equality
$e^{-i\epsilon\ln(z-\epsilon)}=e^{-iz^*}e^{iz}$, 
the asymptotic form of $X_\ell$ is expressed as
\begin{eqnarray}
X_\ell&\rightarrow&
       A^{\rm inc}_\ell e^{-i(z^*-\epsilon\ln\epsilon)}
       +A^{\rm ref}_\ell e^{i(z^*-\epsilon\ln\epsilon)}, 
\nonumber\\
A^{\rm inc}_\ell&=&{1\over2}i^{\ell+1}e^{-i\epsilon\ln\epsilon}
\Bigl[1+\epsilon\left\{P^{(1)}_\ell
                       +i\left(Q^{(1)}_\ell+\ln z\right)\right\}
\cr
&&\qquad
    +\epsilon^2\left\{\left(P^{(2)}_\ell-Q^{(1)}_\ell\ln z\right\}
            +i\left(Q^{(2)}_\ell+P^{(1)}_\ell\ln z\right)\right\}
    +\cdots\Bigr].
\label{eq:Ainlogform}
\end{eqnarray}
Note that
$$
\omega r^*=\omega\left(r+2M\ln{r-2M\over2M}\right)
           =z^*-\epsilon\ln\epsilon,
\label{eq:zrstar}
$$
because of our definition of $z^*$, $z^*=z+\epsilon+\ln(z-\epsilon)$.
The phase factor $e^{-i\epsilon\ln\epsilon}$ of $A^{\rm inc}_\ell$
originates from this definition, but it represents a physical phase
shift due to wave propagation on the curved background.

As one may immediately notice,
the above expression for $A^{\rm inc}_\ell$ 
contains $\ln z$-dependent terms.
Since $A^{\rm inc}_\ell$ should be constant,
$P^{(n)}_\ell$ and $Q^{(n)}_\ell$ should contain appropriate
$\ln z$-dependent terms which exactly cancel the $\ln z$-dependent
terms in the formula (\ref{eq:Ainlogform}).
To be explicit, we must have
\begin{eqnarray}
 P^{(1)}_\ell&=&p^{(1)}_\ell\,,
\cr
 Q^{(1)}_\ell&=&q^{(1)}_\ell-\ln z\,,
\cr
 P^{(2)}_\ell&=&p^{(2)}_\ell+q^{(1)}_\ell\ln z-(\ln z)^2\,,
\cr
 Q^{(2)}_\ell&=&q^{(2)}_\ell-p^{(1)}_\ell\ln z\,,
\label{eq:PQlog}
\end{eqnarray}
where $p^{(n)}_\ell$ and $q^{(n)}_\ell$ are constants.
These relations can be used to check the consistency of
the solution $f^{(n)}$ obtained by integration.
In terms of $p^{(n)}_\ell$ and $q^{(n)}_\ell$,
$A^{\rm inc}_\ell$ is expressed as
\begin{equation}
 A^{\rm inc}_\ell
  ={1\over2}i^{\ell+1}e^{-i\epsilon\ln\epsilon}
    \left[1+\epsilon\left(p^{(1)}_\ell+iq^{(1)}_\ell\right)
          +\epsilon^2\left(p^{(2)}_\ell+iq^{(2)}_\ell\right)
          +\cdots\right].
\label{eq:Ainform}
\end{equation}

Note that the above form of $A^{\rm inc}_\ell$
implies that the so-called tail of radiation, which is due to the curvature
scattering of waves, will contain $\ln v$ terms
as phase shifts in the waveform, but will not give rise to
such terms in the luminosity formula. This supports our previous
argument on the origin of the $\ln v$ terms in the luminosity.
That is, it is not due to the wave propagation effect but due to
some near-zone curvature effect.

%% file: livrev4.texin
\section{Analytic solutions of the homogeneous Teukolsky equation
by means of the series expansion of special functions}
\label{sec:MSTformalism}

In this section, we review a method developed by
Mano, Suzuki and Takasugi~\cite{ref:MST} who
found analytic expressions of the solutions of the homogeneous
Teukolsky equation.
In this method, the exact solutions of the radial Teukolsky equation
(\ref{eq:radTeukolsky}) are expressed in two kinds of series expansions.
One is given by a series of hypergeometric functions
and the other by a series of the Coulomb wave functions. The former is
convergent at horizon and the latter at infinity.
The matching of
these two solutions are done exactly in the overlapping region of
convergence.
They also found that the series expansions are naturally related
to the low frequency expansion. Properties of the analytic
solutions were studied in detail in \cite{ref:MT}.
Thus, the formalism is quite powerful
when dealing with the post-Newtonian expansion, especially at
higher orders.

In many cases when we study the perturbation of a Kerr black hole,
it is more convenient to use the Sasaki-Nakamura equation
since it has the form of a standard wave equation,
similar to the Regge-Wheeler equation.
However, it is not quite suited for investigating analytic
properties of the solution near the horizon.
In contrast, the Mano-Suzuki-Takasugi (MST) formalism
allows us to investigate analytic properties of the solution
near the horizon systematically. Hence, it can be used to
compute the higher order post-Newtonian terms of
the gravitational waves absorbed
into a rotating black hole.

We also note that this method
is the only existing method that can be used to calculate the
gravitational waves emitted to infinity to an arbitrarily high
post-Newtonian order in principle.

\subsection{Angular eigenvalue}

The solutions of the angular equation (\ref{eq:spheroid})
that reduce to the spin-weighted spherical harmonics in the
limit $a\omega\to0$ are called the spin-weighted spheroidal harmonics.
They are the eigen functions of Eq.~(\ref{eq:spheroid}) with
$\lambda$ being the eigenvalues.
The eigenvalues $\lambda$ are necessary for discussions
of the radial Teukolsky equation.
For general spin weight $s$, the spin weighted spheroidal harmonics
obeys 
\begin{eqnarray}
& &\Bigl[{1 \over \sin\theta}{d \over d\theta}
 \Bigl\{\sin\theta {d \over d\theta} \Bigr\}
-a^2\omega^2\sin^2\theta
 -{(m+s\cos\theta)^2 \over \sin^2\theta}
\nonumber\\
& &~~~~~~~~~~~~~~~
-2a\omega s \cos\theta+s+2ma\omega+\lambda\Bigr] 
{}_{s}S_{\ell m}=0. 
\end{eqnarray}

In the post-Newtonian expansion, the parameter $a\omega$ is assumed
to be small. Then, it is straightforward to obtain
a spheroidal harmonic $_{s}S_{\ell m}$ of spin-weight $s$
and its eigenvalue $\lambda$ perturbatively
by the standard method~\cite{ref:PT,ref:TSTS,ref:SSTT}.

It is also possible to obtain the spheroidal harmonics
by expansion in terms of the Jacobi functions~\cite{ref:fackerell}.
In this method, if we calculate numerically, we can obtain them
and their eigenvalues for arbitrary value of $a\omega$.

Here we only show an analytic formula for
the eigenvalue $\lambda$ accurate to $O\left((a\omega)^2\right)$,
which is needed for the calculation of the radial functions.
It is given by
\begin{equation}
\lambda=\lambda_0+a\omega\lambda_1+a^2\omega^2 \lambda_2
+O((a\omega)^3)
\label{eq:pnlambda}
\end{equation}
where
\begin{eqnarray}
&&\lambda_0=\ell(\ell+1)-s(s+1),
\nonumber\\
&&\lambda_1=-2m\left(1+{s^2\over {\ell(\ell+1)}}\right),
\nonumber\\
&&\lambda_2=1+(H(\ell+1)-H(\ell)-1),
\end{eqnarray}
with
\begin{eqnarray}
H(\ell)={{2(\ell^2-m^2)(\ell^2-s^2)^2}\over
{(2\ell-1)\ell^3(2\ell+1)}}.
\end{eqnarray}

\subsection{Horizon solution in series of hypergeometric functions}

As in Sec.~\ref{sec:ReggeWheeler},
we focus on the ingoing-wave function of the radial
Teukolsky equation (\ref{eq:radTeukolsky}).
Since the analysis below is applicable to any spin,
$|s|=0$, $1/2$, $1$, $3/2$ and $2$, we do not specify it
except when it is needed.
Also, the analysis is not restricted to the case $a\omega\ll 1$
unless so stated explicitly.
For general spin weight $s$, the homogeneous Teukolsky equation 
is given by 
\begin{equation}
\Delta^{-s}{d\over dr}\left(\Delta^{s+1}{dR_{\ell m\omega}\over dr}\right)
+\left({{K^2-2is(r-M)K}\over \Delta}
+4is\omega r-\lambda\right)R_{\ell m\omega}=0. 
\label{eq:teukolsky2}
\end{equation}
As before, taking account of the symmetry 
$\bar{R}_{\ell m\omega}=R_{\ell -m -\omega}$,
we may assume $\epsilon=2M\omega>0$ if necessary.

The Teukolsky equation has two regular
singularities at $r=r_{\pm}$, and one irregular singularity at
$r=\infty$. This implies that it cannot be represented in the form
of a single hypergeometric equation. However, if we focus on the
solution near the horizon, it may be approximated by a hypergeometric
equation. This motivates us to consider the solution expressed in terms
of a series of hypergeometric functions.

We define the independent variable $x$ in place of $z$ ($=\omega r$)
as
\begin{equation}
x={{z_+-z}\over \epsilon\kappa}\,,
\end{equation}
where
\[
z_{\pm}=\omega r_{\pm}\,,\quad
\kappa=\sqrt{1-q^2}\,,\quad
q={a\over M}\,.
\]
For later convenience, we also introduce $\tau=(\epsilon-mq)/\kappa$
and $\epsilon_{\pm}=$ $(\epsilon\pm\tau)/2$. 
Taking into account the structure of the singularities at $r=r_{\pm}$,
we put the ingoing-wave Teukolsky function $R_{\ell m\omega}^{\rm in}$ as
\begin{equation}
R^{\rm in}_{\ell m\omega}
=e^{i\epsilon\kappa x}(-x)^{-s-i(\epsilon+\tau)/2}
(1-x)^{i(\epsilon-\tau)/2}p_{\rm in}(x).
\label{eq:hyperrin}
\end{equation}
Then the radial Teukolsky equation becomes
\begin{eqnarray}
&&x(1-x)p_{\rm in}{}''+[1-s-i\epsilon-i\tau-(2-2i\tau)x]p_{\rm in}{}'
\nonumber\\
&&\qquad\qquad+[i\tau(1-i\tau)+\lambda+s(s+1)]p_{\rm in}
\nonumber\\
&&\qquad=2i\epsilon \kappa[-x(1-x)p_{\rm in}{}'
+(1-s+i\epsilon-i\tau)x\,p_{\rm in}]
\nonumber\\
&&\qquad\qquad+[\epsilon^2-i\epsilon \kappa(1-2s)]p_{\rm in}\,,
\label{eq:pin}
\end{eqnarray}
where a prime denotes $d/dx$.
The left-hand-side of Eq.~(\ref{eq:pin}) is in the form
of a hypergeometric equation.
In the limit $\epsilon\to0$, noting Eq.~(\ref{eq:pnlambda}), we find that
a solution which is finite at $x=0$ is given by
\begin{equation}
p_{\rm in}(\epsilon\to0)=F(-\ell-i\tau,\ell+1-i\tau,1-s-i\tau,x).
\label{eq:pine0}
\end{equation}
For a general value of $\epsilon$, Eq.~(\ref{eq:pin}) suggests
that a solution may be expanded in a series of hypergeometric functions
with $\epsilon$ being a kind of an expansion parameter.
This idea was extensively developed by Leaver~\cite{ref:leaver}.
Leaver obtained solutions of the Teukolsky equation expressed
in a series of the Coulomb wave functions. The MST formalism
is an elegant reformulation of the one by Leaver~\cite{ref:leaver}.

The essential point is to introduce the so-called renormalized
angular momentum $\nu$ which is a generalization of $\ell$ to
a non-integer value such that the Teukolsky equation admits
a solution in a convergent series of hypergeometric functions.
Namely, we add the term $[\nu(\nu+1)-\lambda-s(s+1)]\,p_{\rm in}$
to both sides of Eq.~(\ref{eq:pin}) to rewrite it as
\begin{eqnarray}
&&x(1-x)p_{\rm in}{}''+[1-s-i\epsilon-i\tau-(2-2i\tau)x]p_{\rm in}{}'
\nonumber\\
&&\qquad\qquad+[i\tau(1-i\tau)+\nu(\nu+1)]p_{\rm in}
\nonumber\\
&&\qquad=2i\epsilon \kappa[-x(1-x)p_{\rm in}{}'
+(1-s+i\epsilon-i\tau)x\,p_{\rm in}]\nonumber\\
&&\qquad\qquad+[\nu(\nu+1)-\lambda-s(s+1)
+\epsilon^2-i\epsilon \kappa(1-2s)]p_{\rm in}\,.
\label{eq:pinnew}
\end{eqnarray}
Of course, no modification is done to the original equation, and
$\nu$ is just an irrelevant parameter at this stage.
A trick is to consider the right-hand-side of the above equation
as a perturbation, and look for a formal solution specified by
the index $\nu$ in a series expansion form.
Then, only after we obtain the formal solution,
we require that the series should converge and this requirement
determines the value of $\nu$.
Note that, if we take the limit $\epsilon\to0$,
we must have $\nu\to\ell$ (or $\nu\to-\ell-1$), to assure
 $[\nu(\nu+1)-\lambda-s(s+1)]\to0$
and to recover the solution (\ref{eq:pine0})

Let us denote the formal solution specified by
a value of $\nu$ by $p_{\rm in}^{\nu}$. We express it
in the series form,
\begin{eqnarray}
p_{\rm in}^{\nu}&=&\sum_{n=-\infty}^{\infty} a_n\, p_{n+\nu}(x),
\label{eq:pinexpand}
\\
p_{n+\nu}(x)&=&F(n+\nu+1-i\tau,-n-\nu-i\tau;1-s-i\epsilon-i\tau;x).
\end{eqnarray}
Here, the hypergeometric functions $p_{n+\nu}(x)$ satisfy
the recurrence relations~\cite{ref:MST},
\begin{eqnarray}
xp_{n+\nu}&=&-{(n+\nu+1-s-i\epsilon)(n+\nu+1-i\tau)
\over{2(n+\nu+1)(2n+2\nu+1)}}p_{n+\nu+1}
\nonumber\\
&+&{1\over{2}}\left[1+{i\tau(s+i\epsilon)\over{(n+\nu)(n+\nu+1)}}\right]
p_{n+\nu}\nonumber\\
&-&{(n+\nu+s+i\epsilon)(n+\nu+i\tau)\over{2(n+\nu)(2n+2\nu+1)}}p_{n+\nu-1},
\end{eqnarray}
\begin{eqnarray}
x(1-x)p'_{n+\nu}&=&{(n+\nu+i\tau)(n+\nu+1-i\tau)(n+\nu+1-s-i\epsilon)
\over{2(n+\nu+1)(2n+2\nu+1)}}p_{n+\nu+1}\nonumber\\
&+&{1\over{2}}(s+i\epsilon)
\left[1+{i\tau(1-i\tau)\over{(n+\nu)(n+\nu+1)}}\right]
p_{n+\nu}\nonumber\\
&-&{(n+\nu+1-i\tau)(n+\nu+i\tau)(n+\nu+s+i\epsilon)
\over{2(n+\nu)(2n+2\nu+1)}}p_{n+\nu-1},
\end{eqnarray}
Inserting the series (\ref{eq:pinexpand}) into Eq.~(\ref{eq:pinnew})
and using the above recurrence relations,
we obtain a three-term recurrence relation
among the expansion coefficients
$a_n$. It is given by
\begin{equation}
\alpha_n^{\nu} a_{n+1}+\beta_n^{\nu} a_n
+\gamma_n^{\nu} a_{n-1}=0,
\label{eq:recurrence}
\end{equation}
where
\begin{eqnarray}
\alpha_n^{\nu}&=&{i\epsilon \kappa
(n+\nu+1+s+i\epsilon)(n+\nu+1+s-i\epsilon)(n+\nu+1+i\tau)
\over{(n+\nu+1)(2n+2\nu+3)}},
\nonumber\\
\beta_n^{\nu}&=&-\lambda-s(s+1)+(n+\nu)(n+\nu+1)
+\epsilon^2+\epsilon(\epsilon-mq)\nonumber\\
&&+{\epsilon (\epsilon-mq)(s^2+\epsilon^2) \over{(n+\nu)(n+\nu+1)}},
\label{eq:alpbetgam}\\
\gamma_n^{\nu}&=&-{i\epsilon \kappa (n+\nu-s+i\epsilon)
(n+\nu-s-i\epsilon)(n+\nu-i\tau)
\over{(n+\nu)(2n+2\nu-1)}}.\nonumber
\end{eqnarray}
The convergence of the series (\ref{eq:pinexpand}) is determined
by the asymptotic behaviors of the coefficients ${a_n^\nu}$
at $n\to\pm\infty$.
We thus discuss properties of
the three-term recurrence relation (\ref{eq:recurrence}), and
the role of the parameter $\nu$ in detail.

The general solution of the recurrence relation (\ref{eq:recurrence})
is expressed in terms of two linearly independent solutions
$\{f_n^{(1)}\}$ and $\{f_{n}^{(2)}\}$ ($n=\pm1$, $\pm2,\cdots$).
According to the theory of three-term recurrence relations
(\cite{ref:gautschi}, p.31) 
when there exists a pair of solutions which satisfy
\begin{equation}
\lim_{n\rightarrow \infty} {f_n^{(1)}\over f_n^{(2)}}=0\quad
\left(\lim_{n\rightarrow -\infty} {f_n^{(1)}\over f_n^{(2)}}
=0\right),
\end{equation}
then the solution $\{f_n^{(1)}\}$
is called {\it minimal} as $n\to\infty$ ($n\to-\infty$).
Any non-minimal solution is called {\it dominant}.
The minimal solution (either as $n\to\infty$ or as $n\to-\infty$)
is determined uniquely up to an overall normalization factor.

The three-term recurrence relation is closely
related to continued fractions.
We introduce
\begin{equation}
R_n\equiv {a_{n}\over a_{n-1}}\,,\quad
L_n\equiv {a_{n}\over a_{n+1}}\,.
\end{equation}
We can express $R_n$ and $L_n$ in terms of continued fractions as
\begin{eqnarray}
R_n&=&-{\gamma_n^\nu\over {\beta_n^\nu+\alpha_n^\nu R_{n+1}}}
\nonumber\\
&=&-{\gamma_{n}^\nu\over \beta_{n}^\nu-}\,
{\alpha_{n}\gamma_{n+1}^\nu\over \beta_{n+1}^\nu-}\,
{\alpha_{n+1}\gamma_{n+2}^\nu\over \beta_{n+2}^\nu-}\cdots,
\label{eq:Rncont}\\
L_n&=&-{\alpha_n^\nu\over {\beta_n^\nu+\gamma_n^\nu L_{n-1}}}
\nonumber\\
&=&-{\alpha_{n}^\nu\over \beta_{n}^\nu-}\,
{\alpha_{n-1}\gamma_{n}^\nu\over \beta_{n-1}^\nu-}\,
{\alpha_{n-2}\gamma_{n-1}^\nu\over \beta_{n-2}^\nu-}\cdots.
\label{eq:Lncont}
\end{eqnarray}
These expressions for $R_n$ and $L_n$ are valid if
the respective continued fractions converge.
It is proved (\cite{ref:gautschi}, p.31) that the continued fraction
(\ref{eq:Rncont}) converges if and only if
the recurrence relation (\ref{eq:recurrence}) possesses
a minimal solution as $n\to\infty$, and
the same for the continued fraction (\ref{eq:Lncont})
as $n\to-\infty$.

Analysis of the asymptotic behavior of (\ref{eq:recurrence})
shows that, as long as $\nu$ is finite,
there exists a set of two independent solutions
which behave as (e.g., \cite{ref:gautschi}, p.35),
\begin{eqnarray}
\lim_{n\rightarrow\infty}n{a_n^{(1)}\over a_{n-1}^{(1)}}
={i\epsilon\kappa\over 2}\,,
\quad
\lim_{n\rightarrow\infty}{a_n^{(2)}\over n a_{n-1}^{(2)}}
={2i\over \epsilon\kappa}\,,
\quad
\label{pinfty}
\end{eqnarray}
and another set of two independent solutions which behave as
\begin{eqnarray}
\lim_{n\rightarrow -\infty}n{b_n^{(1)}\over b_{n+1}^{(1)}}
=-{i\epsilon\kappa\over 2}\,,
\quad
\lim_{n\rightarrow -\infty}{b_n^{(2)}\over n b_{n+1}^{(2)}}
=-{2i\over \epsilon\kappa}\,.
\label{ninfty}
\end{eqnarray}
Thus $\{a_n^{(1)}\}$ is minimal as $n\to\infty$ and
$\{b_n^{(1)}\}$ is minimal as $n\to-\infty$.

Since the recurrence relation (\ref{eq:recurrence}) possesses
minimal solutions as $n\to\pm\infty$,
the continued fractions on the right-hand-sides
of Eqs.~(\ref{eq:Rncont}) and (\ref{eq:Lncont}) converge
for $a_n=a_n^{(1)}$ and $a_n=b_n^{(1)}$.
In general, however, $a_n^{(1)}$ and $b_n^{(1)}$ do not
coincide.
Here, we use the freedom of $\nu$ to obtain a consistent
solution. Let $\{f_n^\nu\}$ be a sequence which is
minimal for both $n\to\pm\infty$.
We then have expressions for
$f_n^\nu/f_{n-1}^\nu$ and $f_n^\nu/f_{n+1}^\nu$ in terms of
continued fractions as
\begin{eqnarray}
\tilde{R}_n\equiv {f_n\over f_{n-1}}
&=&-{\gamma_{n}^\nu\over \beta_{n}^\nu-}\,
{\alpha_{n}\gamma_{n+1}^\nu\over \beta_{n+1}^\nu-}\,
{\alpha_{n+1}\gamma_{n+2}^\nu\over \beta_{n+2}^\nu-}\cdots,
\label{eq:contiR} \\
\tilde{L}_n\equiv {f_n\over f_{n+1}}
&=&-{\alpha_{n}^\nu\over \beta_{n}^\nu-}\,
{\alpha_{n-1}\gamma_{n}^\nu\over \beta_{n-1}^\nu-}\,
{\alpha_{n-2}\gamma_{n-1}^\nu\over \beta_{n-2}^\nu-}\cdots.
\label{eq:contiL}
\end{eqnarray}
This implies
\begin{equation}
\tilde{R}_n \tilde{L}_{n-1}=1\,.
\label{eq:nudefeq}
\end{equation}
Thus, if we choose $\nu$ such that it satisfies the implicit
equation for $\nu$, Eq.~(\ref{eq:nudefeq}),
for a certain $n$, we obtain a unique minimal solution $\{f_n^\nu\}$
which is valid over the entire range of $n$, $-\infty<n<\infty$,
that is
\begin{eqnarray}
\lim_{n\rightarrow\infty}n\,{f_n^\nu\over f_{n-1}^\nu}
={i\epsilon\kappa\over 2}\,,
\quad
\lim_{n\rightarrow -\infty}n\,{f_n^\nu\over f_{n+1}^\nu}
=-{i\epsilon\kappa\over 2}\,.
\label{pninfty}
\end{eqnarray}
Note that if Eq.~(\ref{eq:nudefeq}) for a certain value of $n$
is satisfied, it is automatically
satisfied for any other value of $n$.

The minimal solution is also important for the convergence of
the series (\ref{eq:pinexpand}).
For the minimal solution $\{f_n^\nu\}$,
together with the properties of the hypergeometric
functions $p_{n+\nu}$ for large $|n|$, we find
\begin{eqnarray}
\lim_{n\rightarrow \infty}
n{f_{n+1}^\nu p_{n+\nu+1}(x)\over f_n^\nu p_{n+\nu}(x)}
&=&
-\lim_{n\rightarrow -\infty}
n{f_{n-1}^\nu p_{n+\nu-1}(x)\over f_n^\nu p_{n+\nu}(x)}
\cr
&=&
{i\epsilon\kappa\over 2}[1-2x+((1-2x)^2-1)^{1/2}].
\end{eqnarray}
Thus the series of hypergeometric functions
(\ref{eq:pinexpand}) converges for all $x$ in the range
$0\geq x>-\infty$ (in fact, for all complex values of $x$
except at $|x|=\infty$), provided the coefficients are
given by the minimal solution.

Instead of Eq.~(\ref{eq:nudefeq}),
we may consider an equivalent but practically more convenient form
of an equation that determines the value of $\nu$.
Dividing Eq.~(\ref{eq:recurrence}) by $a_n$, we find
\begin{equation}
\beta_n^{\nu}+\alpha_n^\nu R_{n+1}+\gamma_n^\nu L_{n-1}=0,
\label{eq:nudefeq2}
\end{equation}
where $R_{n+1}$ and $L_{n+1}$ are those given by the
continued fractions (\ref{eq:contiR}) and (\ref{eq:contiL}),
respectively.
Although the value of $n$ in this equation is arbitrary,
it is convenient to set $n=0$ to solve for $\nu$.

For later use, we need a series expression for
$R^{\rm in}$ with better convergence properties
at large $|x|$.
Using analytic properties of hypergeometric functions,
we have
\begin{equation}
R^{\rm in}=R_0^\nu+R_0^{-\nu-1},
\end{equation}
where
\begin{eqnarray}
R_0^{\nu}&=&e^{i\epsilon \kappa x}(-x)^{-s-{i\over{2}}(\epsilon+\tau)}
(1-x)^{{i\over{2}}(\epsilon+\tau)+\nu}
\nonumber\\
&&
\times\sum_{n=-\infty}
^{\infty}f_n^{\nu}\,{\Gamma(1-s-i\epsilon-i\tau)\Gamma(2n+2\nu+1)
\over{\Gamma(n+\nu+1-i\tau)\Gamma(n+\nu+1-s-i\epsilon)}}
\nonumber\\
&\times&
(1-x)^{n}F(-n-\nu-i\tau,-n-\nu-s-i\epsilon;-2n-2\nu;{1\over{1-x}})\,.
\end{eqnarray}
This expression explicitly exhibits the symmetry of $R_{\rm in}$
under the interchange of $\nu$ and $-\nu-1$. This is a result of
the fact that $\nu(\nu+1)$ is invariant under the interchange
$\nu\leftrightarrow-\nu-1$. Accordingly, the recurrence
relation (\ref{eq:recurrence}) has the structure that
$f_{-n}^{-\nu-1}$ satisfies
the same recurrence relation as $f_n^{\nu}$.

Finally, we note that if $\nu$ is a solution of
Eq.~(\ref{eq:nudefeq}) or (\ref{eq:nudefeq2}),
$\nu+k$ with an arbitrary integer $k$ is also a solution,
since $\nu$ appears only in the combination of $n+\nu$.
Thus, Eq.~(\ref{eq:nudefeq}) or (\ref{eq:nudefeq2})
contains infinite number of roots.
However, not all of these can be used to express a solution we want.
As noted in the earlier part of this subsection,
in order to reproduce the solution
in the limit $\epsilon\rightarrow 0$, Eq.~(\ref{eq:pine0}),
we must have $\nu\rightarrow \ell$ (or $\nu\to-\ell-1$ by symmetry).
We thus impose a constraint on $\nu$ such that
it must continuously approach $\ell$ as $\epsilon\rightarrow 0$.

\subsection{Outer solution in series of Coulomb wave functions}

The solution in series of hypergeometric functions discussed in the
previous subsection is convergent
at any finite value of $r$. However, it does not converge at infinity,
and hence the asymptotic amplitudes, $B^{\rm inc}$ and $B^{\rm ref}$,
cannot be determined from it. To determine the asymptotic amplitudes, 
it is necessary to construct a solution which is valid at infinity
and match the two solutions in a region where both solutions converge.
The solution convergent at infinity was obtained by Leaver 
in series of Coulomb wave functions~\cite{ref:leaver}.
In this subsection, we review Leaver's solution 
based on \cite{ref:MT}. 

In this subsection again, by noting the symmetry 
$\bar{R}_{\ell m\omega}=R_{\ell -m -\omega}$,
we assume $\omega>0$ without loss of generality. 

First, we define a variable $\hat{z}=\omega(r-r_-)=\epsilon\kappa(1-x)$.
Let us denote a Teukolsky function by $R_C$.
We introduce a function $f(\hat z)$ by
\begin{equation}
R_C=\hat{z}^{-1-s}
\left(1-{\epsilon\kappa\over \hat{z}}
\right)^{-s-i(\epsilon+\tau)/2} f(\hat{z})\,.
\end{equation}
Then the Teukolsky equation becomes
\begin{eqnarray}
&&\hat{z}^2 f''+[\hat{z}^2+(2\epsilon+2is)\hat{z}-\lambda-s(s+1)]f
\nonumber\\
&&\qquad
=\epsilon\kappa \hat{z} (f''+f)+\epsilon\kappa(s-1+2i\epsilon)f'
\nonumber\\
&&\qquad\quad
-{\epsilon\over \hat{z}}(\kappa-i(\epsilon-mq))(s-1+i\epsilon)f
\nonumber\\
&&\qquad\quad
+[-2\epsilon^2+\epsilon mq+\kappa(\epsilon^2+i\epsilon s)]f\,.
\label{eq:fnueq}
\end{eqnarray}
We see that the right-hand-side is explicitly of 
$O(\epsilon)$ and the left-hand-side is in the form of
the Coulomb wave equation. Therefore, in the limit $\epsilon\to0$,
we obtain a solution,
\begin{equation}
f(\hat{z})=F_\ell(-is-\epsilon, \hat{z}),
\end{equation}
where $F_L(\eta,\hat{z})$ is a Coulomb wave function given by 
\begin{eqnarray}
F_{L}(\eta,\hat{z})=e^{-i\hat{z}}2^L\hat{z}^{L+1}
{\Gamma(L+1-i\eta)\over{\Gamma(2L+2)}}
\Phi(L+1-i\eta,2L+2;2i\hat{z}),
\label{eq:defcoulomb}
\end{eqnarray}
and $\Phi$ is the regular confluent hypergeometric function
function (see \cite{ref:handbook}, section 13) which is regular at 
$\hat{z}=0$.

In the same spirit as in the previous subsection, 
we introduce the renormalized angular momentum $\nu$.
That is, we add $[\lambda+s(s+1)-\nu(\nu+1)]f(\hat{z})$ to both sides
of Eq.~(\ref{eq:fnueq}) to rewrite it as 
\begin{eqnarray}
&&\hat{z}^2 f''+[\hat{z}^2+(2\epsilon+2is)\hat{z}-\nu(\nu+1)]f
\nonumber\\
&&\qquad=\epsilon\kappa \hat{z} (f''+f)
+\epsilon\kappa(s-1+2i\epsilon_+)f'
-{\epsilon\over \hat{z}}(\kappa-i(\epsilon-mq))(s-1+i\epsilon)f
\nonumber\\
&&\qquad\quad
+[-\nu(\nu+1)+\lambda+s(s+1)
-2\epsilon^2+\epsilon mq+\kappa(\epsilon^2+i\epsilon s)]f\,.
\label{eq:fnueq2}
\end{eqnarray}
We denote the formal solution specified by the index $\nu$ by 
$f_\nu(\hat{z})$, and
expand it in terms of the Coulomb wave functions as
\begin{equation}
f_\nu=\sum_{n=-\infty}^{\infty}
(-i)^n{(\nu+1+s-i\epsilon)_n\over (\nu+1-s+i\epsilon)_n}\,
b_n \,F_{n+\nu}(-is-\epsilon, \hat{z})\,,
\label{eq:expandfnu}
\end{equation}
where $(a)_n=\Gamma(a+n)/\Gamma(a)$.
Then, using the recurrence relations among $F_{n+\nu}$,
\begin{eqnarray}
{1\over{z}}F_{n+\nu}&=&{(n+\nu+1+s-i\epsilon)
\over{(n+\nu+1)(2n+2\nu+1)}}F_{n+\nu+1}
+{is+\epsilon\over{(n+\nu)(n+\nu+1)}}F_{n+\nu}\nonumber\\
&+&{(n+\nu-s+i\epsilon)\over{(n+\nu)(2n+2\nu+1)}}F_{n+\nu-1},
\\
F'_{n+\nu}&=&-{(n+\nu)(n+\nu+1+s-i\epsilon)
\over{(n+\nu+1)(2n+2\nu+1)}}F_{n+\nu+1}
+{is+\epsilon\over{(n+\nu)(n+\nu+1)}}F_{n+\nu}\nonumber\\
&+&{(n+\nu+1)(n+\nu-s+i\epsilon)\over{(n+\nu)(2n+2\nu+1)}}F_{n+\nu-1}\,,
\end{eqnarray}
we can derive the recurrence relation among $b_n$.
The result turns out to be identical to the one given 
by Eq.~(\ref{eq:recurrence}) for $a_n$. We mention that
the extra factor $(\nu+1+s-i\epsilon)_n/(\nu+1-s+i\epsilon)_n$
in Eq.~(\ref{eq:expandfnu}) is introduced to make the recurrence
relation exactly identical to Eq.~(\ref{eq:recurrence}).

The fact that we have the same recurrence relation as 
Eq.~(\ref{eq:recurrence}) implies that if we choose the parameter 
$\nu$ in Eq.~(\ref{eq:expandfnu}) to be the same 
as the one given by a solution of Eq.~(\ref{eq:nudefeq})
 or (\ref{eq:nudefeq2}), 
the sequence $\{f_n^\nu\}$ is also the solution for $\{b_n\}$
which is minimal for both $n\rightarrow\pm\infty$. 
Let us set
$$
g_n^\nu=(-i)^n{(\nu+1+s-i\epsilon)_n\over (\nu+1-s+i\epsilon)_n}\,f_n^\nu\,.
$$
By choosing $\nu$ as stated above, 
we have the asymptotic value for the ratio of two successive terms 
of $g_n^\nu$ as
\begin{equation}
\lim_{n\rightarrow\infty}n\,{g_n^\nu\over g_{n-1}^\nu}
=\lim_{n\rightarrow -\infty}n\,{g_n^\nu\over g_{n+1}^\nu}
={\epsilon\kappa\over 2}\,.
\end{equation}
Using an asymptotic property of the Coulomb wave functions,
we have
\begin{equation}
\lim_{n\rightarrow \infty}
{g_n^\nu F_{n+\nu}(z)\over g_{n-1}^\nu F_{n+\nu-1}(z)}
=\lim_{n\rightarrow -\infty}
{g_n^\nu F_{n+\nu}(z)\over g_{n-1}^\nu F_{n+\nu+1}(z)}
={\epsilon\kappa\over z}\,.
\end{equation}
We thus find that the series (\ref{eq:expandfnu})
converges at $\hat{z}>\epsilon\kappa$
or equivalently $r>r_+$.

The fact that we can use the same $\nu$ as in the case of 
hypergeometric functions 
to obtain the convergence of the series of the Coulomb wave functions
is crucial to match the horizon and outer solutions. 

Here, we note 
an analytic property of the confluent hypergeometric function 
(\cite{ref:HTF},p.259),
\begin{equation}
\Phi(a,c;x)={\Gamma(c)\over \Gamma(c-a)}e^{ia\pi}\Psi(a,c;x)
+{\Gamma(c)\over \Gamma(a)}e^{i\pi(a-c)}e^x\Psi(c-a,c;-x),
\end{equation}
where $\Psi$ is the irregular confluent hypergeometric function,
and ${\rm Im}(x)>0$ is assumed.
Using this with the identifications,
\begin{eqnarray}
a&=&n+\nu+1-s+i\epsilon\,, \nonumber\\
c&=&2(n+\nu+1)\,,\nonumber\\
x&=&2i\hat z\,,
\nonumber
\end{eqnarray}
we can rewrite $R_C^\nu$ (for $\omega>0$) as
\begin{equation}
R_C^\nu=R_+^\nu+R_-^\nu,
\end{equation}
where
\begin{eqnarray}
R_{+}^{\nu}&=& 2^{\nu}e^{-\pi \epsilon}e^{i\pi(\nu+1-s)}
\frac{\Gamma(\nu+1-s+i\epsilon)}{\Gamma(\nu+1+s-i\epsilon)}
e^{-i\hat{z}}\hat{z}^{\nu+i\epsilon_+}
(\hat{z}-\epsilon\kappa)^{-s-i\epsilon_+}\nonumber\\
&\times&  \sum_{n=-\infty}^{\infty}i^n
f_n^{\nu}(2\hat{z})^n
\Psi(n+\nu+1-s+i\epsilon,2n+2\nu+2;2i\hat{z}),\nonumber\\
&&\\
R_{-}^{\nu}&=& 2^{\nu}e^{-\pi \epsilon}e^{-i\pi(\nu+1+s)}
e^{i\hat{z}}\hat{z}^{\nu+i\epsilon_+}(\hat{z}-\epsilon\kappa)^{-s-i\epsilon_+}
\nonumber\\
&\times&\sum_{n=-\infty}^{\infty}i^n
\frac{(\nu+1+s-i\epsilon)_n}{(\nu+1-s+i\epsilon)_n}
f_n^{\nu}(2\hat{z})^n \nonumber\\
&&\times
\Psi(n+\nu+1+s-i\epsilon,2n+2\nu+2;-2i\hat{z}).
\label{eq:RupMST}
\end{eqnarray}
By noting an asymptotic behavior of $\Psi(a,c;x)$ at large $|x|$,
\begin{equation}
\Psi(a,c;x)\rightarrow x^{-a}
\quad\quad \mbox{as $|x|\rightarrow \infty$},
\end{equation}
we find 
\begin{eqnarray}
&&R_{+}^\nu=A_{+}^{\nu} z^{-1}e^{-i(z+\epsilon \ln z)},\\
&&R_-^\nu=A_{-}^{\nu} z^{-1-2s} e^{i(z+\epsilon \ln z)},
\end{eqnarray}
where
\begin{eqnarray}
&&A_{+}^\nu=e^{-{\pi\over 2}\epsilon}e^{{\pi\over 2}i(\nu+1-s)}
2^{-1+s-i\epsilon}{\Gamma(\nu+1-s+i\epsilon)\over 
\Gamma(\nu+1+s-i\epsilon)}\sum_{n=-\infty}^{+\infty}f_n^\nu,\\
&&A_{-}^\nu=2^{-1-s+i\epsilon}e^{-{\pi\over 2}i(\nu+1+s)}e^{-{\pi\over 2}\epsilon}
\sum_{n=-\infty}^{+\infty}(-1)^n{(\nu+1+s-i\epsilon)_n\over 
(\nu+1-s+i\epsilon)_n}f_n^\nu.
\end{eqnarray}
We can see that the functions $R_{+}^{\nu}$ and $R_{-}^{\nu}$ are
incoming-wave and outgoing-wave solutions at infinity, respectively.
In particular, we have the upgoing solution, defined for $s=-2$ by the
asymptotic behavior~(\ref{eq:Rup}), expressed 
in terms of a series of Coulomb wave functions as
\begin{equation}
R^{\rm up}=R_{-}^{\nu}\,.
\label{Rupsol}
\end{equation}

\subsection{Matching of horizon and outer solutions}

Now, we match the two types of solutions $R_0^\nu$ and $R_C^\nu$.
Note that both of them are convergent in a very large region of $r$,
namely for $\epsilon\kappa<\hat z<\infty$.
We see that both solutions behave a
$\hat{z}^\nu$ multiplied by a single-valued function of $\hat z$
for large $|\hat z|$.
Thus, the analytic properties of ${R}_{0}^{\nu}$  and ${R}_{C}^{\nu}$
are the same, which implies that these two are identical up to
a constant multiple.
Therefore, we set
\begin{equation}
{R}_{0}^{\nu} =K_{\nu}{R}_{C}^{\nu}.
\end{equation}

In the region $\epsilon\kappa<\hat z<\infty$,
we may expand both solutions in powers of
$\hat{z}$ except for analytically non-trivial factors. We have
\begin{eqnarray}
R_0^\nu&=&e^{i\epsilon\kappa} e^{-i\hat{z}}(\epsilon\kappa)^{-\nu-i\epsilon_+}
\hat{z}^{\nu+i\epsilon_+}
\left({\hat{z}\over \epsilon\kappa}-1\right)^{-s-i\epsilon_+}
\sum_{n=-\infty}^{\infty}\sum_{j=0}^\infty
C_{n,j}\hat{z}^{n-j},
\nonumber\\
&=&
e^{i\epsilon\kappa} e^{-i\hat{z}}(\epsilon\kappa)^{-\nu-i\epsilon_+}
\hat{z}^{\nu+i\epsilon_+}
\left({\hat{z}\over \epsilon\kappa}-1\right)^{-s-i\epsilon_+}
\sum_{k=-\infty}^{\infty}\sum_{n=k}^\infty
C_{n,n-k}\hat{z}^{k},\nonumber\\
\\
R_C^\nu&=&e^{-i\hat{z}}2^\nu(\epsilon\kappa)^{-s-i\epsilon_+}
\hat{z}^{\nu+i\epsilon_+}
\left({\hat{z}\over \epsilon\kappa}-1\right)^{-s-i\epsilon_+}
\sum_{n=-\infty}^{\infty}\sum_{j=0}^\infty
D_{n,j}\hat{z}^{n+j},
\nonumber\\
&=&
e^{-i\hat{z}}2^\nu(\epsilon\kappa)^{-s-i\epsilon_+}
\hat{z}^{\nu+i\epsilon_+}
\left({\hat{z}\over \epsilon\kappa}-1\right)^{-s-i\epsilon_+}
\sum_{k=-\infty}^{\infty}\sum_{n=-\infty}^k
D_{n,k-n}\hat{z}^{k},\nonumber\\
\end{eqnarray}
where
\begin{eqnarray}
C_{n,j}&=&{\Gamma(1-s-2i\epsilon_+)\Gamma(2n+2\nu+1)\over 
\Gamma(n+\nu+1-i\tau)\Gamma(n+\nu+1-s-i\epsilon)}
\nonumber\\
&&\times
{{(-n-\nu-i\tau)_j(-n-\nu-s-i\epsilon)_j}\over (-2n-2\nu)_j(j!)}
(\epsilon\kappa)^{-n+j}f_n,
\\
D_{n,j}&=&(-1)^n (2i)^{n+j}
{\Gamma(n+\nu+1-s+i\epsilon)\over \Gamma(2n+2\nu+2)}
{(\nu+1+s-i\epsilon)_n\over (\nu+1-s+i\epsilon)_n}
\nonumber\\
&&\times
{(n+\nu+1-s+i\epsilon)_j\over (2n+2\nu+2)_j(j!)} f_n.
\end{eqnarray}

Then, by comparing each integer power of $\hat{z}$ in the summation, 
in the region $\epsilon\kappa\ll \hat{z} < \infty$, and using 
the formula $\Gamma(z)\Gamma(1-z)$ $=\pi/\sin\pi z$, we find
\begin{eqnarray}
{K}_{\nu}&=&
e^{i\epsilon\kappa}(\epsilon\kappa)^{s-\nu}2^{-\nu}
\left(\sum_{n=-\infty}^{r}D_{n,r-n}\right)^{-1}
\left(\sum_{n=r}^{\infty}C_{n,n-r}\right)
\nonumber\\
&=&
\frac{e^{i\epsilon\kappa}(2\epsilon \kappa )^{s-\nu-r}2^{-s}i^{r}
\Gamma(1-s-2i\epsilon_+)\Gamma(r+2\nu+2)}
{\Gamma(r+\nu+1-s+i\epsilon)
\Gamma(r+\nu+1+i\tau)\Gamma(r+\nu+1+s+i\epsilon)}
\nonumber\\
&&\times \left ( \sum_{n=r}^{\infty}
(-1)^n\, \frac{\Gamma(n+r+2\nu+1)}{(n-r)!}
\frac{\Gamma(n+\nu+1+s+i\epsilon)}{\Gamma(n+\nu+1-s-i\epsilon)}
\frac{\Gamma(n+\nu+1+i\tau)}{\Gamma(n+\nu+1-i\tau)}
\,f_n^{\nu}\right)
\nonumber\\
&&\times \left(\sum_{n=-\infty}^{r}
\frac{(-1)^n}{(r-n)!
(r+2\nu+2)_n}\frac{(\nu+1+s-i\epsilon)_n}{(\nu+1-s+i\epsilon)_n}
f_n^{\nu}\right)^{-1},
\label{eq:Knu}
\end{eqnarray}
where $r$ can be any integer, 
and the factor ${K}_{\nu}$ should be 
independent of the choice of $r$. 
Although this fact is not manifest from Eq.~(\ref{eq:Knu}),
we can check it numerically, or analytically 
by expanding it in terms of $\epsilon$. 

We thus have two expressions for the ingoing-wave function
$R^{\rm in}$.
One is given by Eq.~(\ref{eq:hyperrin}) with $p_{\rm in}^\nu$ 
expressed in terms of a series of hypergeometric functions
as given by Eq.~(\ref{eq:pinexpand}),
which converges everywhere except at $r=\infty$.
The other is expressed in terms of a series of Coulomb wave functions
given by
\begin{equation}
R^{{\rm in}}=K_{\nu}{R}_{C}^{\nu}+
K_{-\nu-1}{R}_{C}^{-\nu-1},
\end{equation}
which converges at $r>r_+$, including $r=\infty$. Combining these
two, we have a complete analytic solution for
the ingoing-wave function.

Now we can obtain analytic expressions for
the asymptotic amplitudes of $R^{\rm in}$,
$B^{\rm trans}$, $B^{\rm inc}$ and $B^{\rm ref}$.
By investigating the asymptotic behaviors of
the solution at $r\rightarrow \infty$ and $r\rightarrow r_+$,
they are found to be
\begin{eqnarray}
B^{\rm trans}&=&\left(\frac{\epsilon\kappa}{\omega}\right)^{2s}
e^{i\epsilon_{+}\ln\kappa}\sum_{n=-\infty}^{\infty}f_{n}^{\nu}\,.
\\
B^{\rm inc}
&=&\omega^{-1}\left[{K}_{\nu}-
ie^{-i\pi\nu} \frac{\sin \pi(\nu-s+i\epsilon)}
{\sin \pi(\nu+s-i\epsilon)}
{K}_{-\nu-1}\right]A_{+}^{\nu} e^{-i\epsilon\ln\epsilon},
\\
B^{\rm ref}
&=&\omega^{-1-2s}\left[{K}_{\nu}
+ie^{i\pi\nu} {K}_{-\nu-1}\right]A_{-}^{\nu}
e^{i\epsilon\ln\epsilon}.
\end{eqnarray}

Incidentally, since we have
the upgoing solution in the outer region~(\ref{Rupsol}),
it is straightforward to obtain 
the asymptotic outgoing amplitude at infinity
$C^{\rm trans}$ from Eq.~(\ref{eq:RupMST}).
We find
\begin{eqnarray}
C^{\rm trans}
=\omega^{-1-2s}e^{i\epsilon\ln\epsilon}A_{-}^\nu.
\end{eqnarray}

\subsection{Low frequency expansion of hypergeometric expansion}

So far, we have considered exact solutions of the Teukolsky equation. 
Now let us consider their low frequency approximation
and determine the value of $\nu$. We solve Eq.~(\ref{eq:nudefeq2})
with $n=0$,
\begin{equation}
\beta_0^{\nu}+\alpha_0^\nu R_{1}+\gamma_{0}^\nu L_{-1}=0,
\label{eq:nudefeq3}
\end{equation}
with a requirement that $\nu\rightarrow \ell$ 
as $\epsilon\rightarrow 0$.

To solve Eq.~(\ref{eq:nudefeq3}), we first note the 
following. Unless the value of $\nu$ is such that
the denominator in the expression of $\alpha_n^\nu$
or $\gamma_n^\nu$ happens to vanish or $\beta_n^\nu$ happens to
vanish in the limit $\epsilon\to0$, we have
$\alpha_n^\nu=O(\epsilon)$, $\gamma_n^\nu=O(\epsilon)$ and
$\beta_n^\nu=O(1)$. Also, from the asymptotic behavior of
the minimal solution $f_n^\nu$ as $n\to\pm\infty$ given by
Eq.~(\ref{pninfty}), we have $R_n(\nu)=O(\epsilon)$
and $L_{-n}(\nu)=O(\epsilon)$ for sufficiently large $n$.
Thus, except for exceptional cases
mentioned above, the order of $a_n^\nu$ in
$\epsilon$ increases as $|n|$ increases. That is,
the series solution naturally gives the post-Minkowski expansion.

First let us consider the case of $R_n(\nu)$ for $n>0$.
It is easily seen that $\alpha_n^\nu=O(\epsilon)$,
$\gamma_n^\nu=O(\epsilon)$ and $\beta_n^\nu=O(1)$ for all $n>0$.
Therefore, we have $R_n(\nu)=O(\epsilon)$ for all $n>0$.

On the other hand, for $n<0$, the order of $L_{-n}(\nu)$
behaves irregularly for certain values of $n$. 
For the moment, let us assume that $L_{-1}(\nu)=O(\epsilon)$.
We see from Eqs.~(\ref{eq:alpbetgam}) that $\alpha_0^\nu=O(\epsilon)$,
$\gamma_0^\nu=O(\epsilon)$, since $\nu=\ell+O(\epsilon)$.
Then, Eq.~(\ref{eq:nudefeq3}) implies $\beta_0^\nu=O(\epsilon^2)$.
Using the expansion of $\lambda$ given by Eq.~(\ref{eq:pnlambda}), we
then find $\nu=\ell+O(\epsilon^2)$, i.e., there is no term of
$O(\epsilon)$ in $\nu$.
With this estimate of $\nu$, we see from Eq.~(\ref{eq:Lncont}) 
that $L_{-1}(\nu)=O(\epsilon)$ is justified
if $L_{-2}(\nu)$ is of order unity or smaller.

The general behavior of the order of $L_{-n}(\nu)$ in $\epsilon$
for general values of $s$ is rather complicated. However, if we
assume $s$ to be a non-integer and $\ell\geq|s|$, and
$\tau=(\epsilon-mq)/\kappa=O(1)$, it is relatively
easily studied. With the assumption that $\nu=\ell+O(\epsilon^2)$,
we find there are three exceptional cases:
For $n=-2\ell-1$, we have $\alpha_n=O(\epsilon)$,
$\beta=O(\epsilon^2)$ and $\gamma_n=O(\epsilon)$.
For $n=-\ell-1$, we have
$\alpha_n^\nu=O(1/\epsilon)$ and $\beta_n^\nu=O(1/\epsilon)$ 
and $\gamma_n=O(\epsilon)$.
And for $n=-\ell$, we have $\alpha_n=O(\epsilon)$, 
$\beta_n=O(1/\epsilon)$ and $\gamma_n=O(1/\epsilon)$.
These implies consecutively that
$L_{-2\ell-1}(\nu)=O(1/\epsilon)$, $L_{-\ell-1}(\nu)=O(1)$
and $L_{-\ell}(\nu)=O(\epsilon^2)$.
To summarize, we have
\begin{eqnarray}
 && R_n(\nu)={f_{n}^\nu\over f_{n-1}^\nu}=O(\epsilon)
\quad\hbox{\rm for all}~n>0,
\nonumber\\
 && L_{-\ell}(\nu)={f_{-\ell}^\nu\over f_{-\ell+1}^\nu}=O(\epsilon^2),
\quad
L_{-\ell-1}(\nu)={f_{-\ell-1}^\nu\over f_{-\ell}^\nu}=O(1),
\nonumber\\
&& L_{-2\ell-1}(\nu)={f_{-2\ell-1}^\nu\over f_{-2\ell}^\nu}=O(1/\epsilon),
\nonumber\\
 && L_{n}(\nu)={f_n^\nu\over f_{n+1}^\nu}=O(\epsilon)
\quad\hbox{\rm for all the other}~n<0.
\end{eqnarray}
With these results, we can calculate the value of $\nu$ to $O(\epsilon)$,
which is given by 
\begin{eqnarray}
\label{eq:nusol}
\nu&=& \ell+{1\over{2\ell+1}}\left[-2-{s^2\over{\ell(\ell+1)}}
+{[(\ell+1)^2-s^2]^2
\over{(2\ell+1)(2\ell+2)(2\ell+3)}}
\right.
\nonumber\\
&&
\left.
-{(\ell^2-s^2)^2\over{(2\ell-1)2\ell(2\ell+1)}}\right]
\epsilon^2+O(\epsilon^3).
\end{eqnarray}
Now one can take the limit of an integer value of $s$.
In particular, the above holds also for $s=0$.
Interestingly, $\nu$ is found to be independent of the azimuthal
eigenvalue $m$ to $O(\epsilon^2)$. 

The post-Minkowski expansion of homogeneous
Teukolsky functions can be obtained with arbitrary accuracy by solving
Eq.~(\ref{eq:recurrence}) to a desired order and by summing
up the terms to a sufficiently large $|n|$.
First few terms of the
coefficients $f_n^{\nu}$ are explicitly given in \cite{ref:MST}. 
A calculation up to a much higher order in $O(\epsilon)$ was performed 
in \cite{ref:TMT}, in which the black hole absorption
of gravitational waves was calculated to $O(v^8)$
beyond the lowest order.

%% file: livrev5.texin
\section{Gravitational waves from a particle orbiting a black hole} 
\label{sec:luminosity}

Based on the ingoing wave functions discussed in the 
previous two sections, we can derive gravitational wave energy and 
angular momentum flux emitted to infinity. 
The formula for the energy and the angular momentum luminosity 
to infinity are given by Eqs.~(\ref{eq:Eflux}) and (\ref{eq:Jflux}).
Since most of the calculations are very long, we only show
the final results. In \cite{ref:supplement}, some details 
of the calculations are summarized. 
We define the post-Newtonian expansion parameter by 
$x\equiv(M\Omega_\varphi)^{1/3}$ where $M$ is the mass of the black hole,
$\Omega_\varphi$ is the orbital angular frequency of the particle. 
Since the parameter $x$ is directly related to
the observable frequency, 
this result can be compared with the results by another method easily. 

\subsection{Circular orbit around a Schwarzschild black hole}
First, we show gravitational waves luminosity emitted by a particle 
in circular orbit around the Schwarzschild black hole~\cite{ref:TS,ref:TTS}.
In this case, $\Omega_\varphi$ is given, 
in the Schwarzschild coordinate, by $\Omega_\varphi=(M/r_0^3)^{1/2}$ 
$\equiv \Omega_c$, where $r_0$ is the orbital radius. 
The luminosity to $O(x^{11})$ is given by
\begin{eqnarray}
\left\langle{dE\over dt}\right\rangle&=&\left({dE\over dt}\right)_N
\nonumber\\
&\times&\Biggl[
1 - {1247\over 336}\,x^2 + 4\,\pi\,{x^3} -{44711\over 9072}\,{x^4} 
-{8191\,\pi \over 672}\,{x^5} 
\nonumber\\
&&+\left( {{6643739519}\over {69854400}} -{{1712\,\gamma }\over {105}}
 + {{16\,{{\pi }^2}}\over 3} -{{3424\,\ln 2}\over {105}} 
- {{1712\,\ln x}\over {105}} \right)\,{x^6}
\nonumber\\
&&-{16285\,\pi \over {504}}\,{x^7} 
\nonumber\\
&&+\left( -{{323105549467}\over {3178375200}} + 
   {{232597\,\gamma }\over {4410}}  -{{1369\,{{\pi }^2}}\over{126}}
\right.
\nonumber\\
&&\qquad
\left. + {{39931\,\ln 2}\over {294}} 
   - {{47385\,\ln 3}\over {1568}} + {{232597\,\ln x}\over {4410}}
          \right)\,{x^8}
\nonumber\\
&&+  \left( {{265978667519\,\pi }\over {745113600}}
  -{{6848\,{\gamma}\,\pi }\over {105}} 
  -{{13696\,\pi \,\ln 2}\over {105}} 
  -{{6848\,\pi\,{\ln x}}\over {105}} \right) x^9 
\nonumber\\
&& + \left( -{{2500861660823683}\over {2831932303200}} 
  + {{916628467\,{\gamma}}\over {7858620}} 
  - {{424223\,{{\pi }^2}}\over {6804}}\right. 
\nonumber\\
&&\qquad
 \left. - {{83217611\,\ln 2}\over {1122660}} 
  + {{47385\,\ln 3}\over {196}}
  + {{916628467\,{\ln x}}\over {7858620}}\right) x^{10}
\nonumber\\
&&+\left( {{8399309750401\,\pi }\over {101708006400}} 
  + {{177293\,{\gamma}\,\pi }\over {1176}}\right.
\nonumber\\
&&\qquad
 \left. + {{8521283\,\pi \,\ln 2}\over {17640}} 
  - {{142155\,\pi \,\ln 3}\over {784}}
  + {{177293\,\pi\,{\ln x}}\over {1176}}\right) x^{11} \Biggr],
\label{eq:dEdt11}
\end{eqnarray}
where $(dE/dt)_N$ is the Newtonian quadrupole luminosity given by
\begin{equation}
\left({dE\over dt}\right)_{\rm N}={32\mu^2M^3\over5r_0^5}
={32\over5}\left({\mu\over M}\right)^2x^{10}.
\label{eq:Enewton}
\end{equation}
This is the 5.5PN formula beyond the lowest, 
Newtonian quadrapole formula. 
We can find that our result agrees with the standard post-Newtonian 
results up to $O(x^5)$~\cite{ref:BDIWW}
in the limit  $\mu/M$ $\ll 1$. 

\subsection{Circular orbit on the equatorial plane around a Kerr black
hole}

Next, we consider a particle 
in circular orbit on the equatorial plane 
around a Kerr black hole~\cite{ref:TSTS}. 
In this case, the orbital angular frequency $\Omega_\varphi$ is given by 
\begin{eqnarray}
\Omega_\varphi=
\Omega_c\left[1-q v^3+q^2 v^6+O(v^9)\right],
\end{eqnarray}
where $\Omega_c$ is the orbital angular frequency 
of the circular orbit in Schwarzschild case, 
$v=(M/r_0)^{1/2}$, $q=a/M$, and $r_0$ is the orbital radius
in the Boyer-Lindquist coordinate. 
The effect of the angular momentum of the black hole is 
given by the corrections depending on the parameter $q$. 
Here, $q$ is arbitrary as long as $|q|<1$. 
The luminosity is given  up to $O(x^8)$ (4PN order) by 
\begin{eqnarray}
\left\langle{dE \over dt}\right\rangle
 &=&\left({dE\over dt}\right)_{\rm N}
\left( 1 + \left(\hbox{$q$-independent terms}\right)
 - {{11\,q}\over 4}\,{x^3} + {{33\,{q^2}}\over {16}} \,{x^4} 
\right.\nonumber\\
&-& {{59\,q}\over {16}}\,{x^5} 
+\left( - {{65\,\pi \,q}\over 6} + 
         {{611\,{q^2}}\over {504}}\right)\,{x^6} 
\nonumber\\
&+& \left( {{162035\,q}\over {3888}} +{{65\,\pi \,{q^2}}\over 8} 
          - {{71\,{q^3}}\over {24}} \right) \,{x^7} 
\nonumber\\
&+&\left.
 \left({{359\,\pi \,q}\over {14}} + {{22667\,{q^2}}\over {4536}} 
        + {{17\,{q^4}}\over {16}}\right)\,{x^8}\right). 
\end{eqnarray}

\subsection{Slightly eccentric orbit around 
a Schwarzschild black hole}

Next, we consider a particle 
in slightly eccentric orbit on the equatorial plane 
around a Schwarzschild black hole (\cite{ref:supplement},Sec.7). 
We define $r_0$ as the minimum of radial potential $R(r)/r^4$. 
We also define an eccentricity parameter $e$ from the maximum 
radius of the orbit $r_{\rm max}$, which is given by 
$r_{\rm max}=r_0(1+e)$. 
These conditions are explicitly given by 
\begin{eqnarray}
{\partial(R/r^4)\over \partial r}\bigg|_{r=r_0}=0,
\quad\mbox{and}\quad R(r_0(1+e))=0.
\end{eqnarray}
We assume $e\ll 1$. 
In this case, $\Omega_\varphi$ is given to $O(e^2)$ by 
\begin{eqnarray}
\Omega_\varphi=\Omega_c\left[1-f(v)e^2\right],
\quad\quad
f(v)={3(1-3v^2)(1-8v^2)\over 2(1-2v^2)(1-6v^2)},
\end{eqnarray}
where $\Omega_c=(M/r_0^3)^{1/2}$ is the orbital 
angular frequency in the circular orbit case.
We show the energy and angular momentum luminosity
which are accurate to $O(e^2)$ and to $O(x^8)$ beyond
the Newtonian order. They are given by 
\begin{eqnarray}
\left\langle{dE \over dt}\right\rangle
&=&\left({dE\over dt}\right)_{\rm N}
\Biggl[1 + \left(\hbox{\rm $e$-independent terms}\right)
\nonumber\\
&&+ 
  {e^2}\,\left( {{157}\over {24}} - 
     {{6781\,{x^2}}\over {168}} + 
     {{2335\,\pi \,{x^3}}\over {48}} - 
     {{14929\,{x^4}}\over {189}} - 
     {{773\,\pi \,{x^5}}\over 3} \right.
\nonumber\\
&&+{{156066596771\,{x^6}}\over {69854400}} 
-{{106144\,{\gamma}\,{x^6}}\over {315}} 
+{{992\,{{\pi }^2}\,{x^6}}\over 9} 
\nonumber\\
&&- {{80464\,{x^6}\,\ln 2}\over {315}} 
-{{234009\,{x^6}\,\ln 3}\over {560}} 
-{{106144\,{x^6}\,\ln x}\over {315}} 
\nonumber\\
&&-{{32443727\,\pi \,{x^7}}\over {48384}} 
-{{3045355111074427\,{x^8}}\over {671272842240}} 
+{{507208\,{\gamma}\,{x^8}}\over {245}} 
\nonumber\\
&&
-{{31271\,{{\pi }^2}\,{x^8}}\over {63}} 
-{{151336\,{x^8}\,\ln 2}\over {441}} 
+{{12887991\,{x^8}\,\ln 3}\over {3920}} 
\nonumber\\
&&\left.
+{{507208\,{x^8}\,\ln x}\over {245}} 
\right) \Biggr],
\label{eq:esdedtx}
\end{eqnarray}
and
\begin{eqnarray}
\left\langle{dJ_z \over dt}\right\rangle
&=&\left({dJ\over dt}\right)_{\rm N}
\Biggl[1+\left(\hbox{\rm $e$-independent terms}\right)
\nonumber\\
&&+{e^2}\,\left( {{23}\over 8} 
- {{3259\,{x^2}}\over {168}} 
+{{209\,\pi \,{x^3}}\over 8} 
-{{1041349\,{x^4}}\over {18144}} 
-{{785\,\pi \,{x^5}}\over 6} \right.
\nonumber\\
&&+{{91721955203\,{x^6}}\over {69854400}} 
-{{41623\,{\gamma}\,{x^6}}\over {210}} 
+{{389\,{{\pi }^2}\,{x^6}}\over 6} 
-{{24503\,{x^6}\,\ln 2}\over {210}} 
\nonumber\\
&&
-{{78003\,{x^6}\,\ln 3}\over {280}} 
-{{41623\,{x^6}\,\ln x}\over {210}} 
-{{91565\,\pi \,{x^7}}\over {168}} 
\nonumber\\
&&
-{{105114325363\,{x^8}}\over {72648576}} 
+{{696923\,{\gamma}\,{x^8}}\over {630}} 
-{{4387\,\pi^2\,{x^8}}\over {18}} 
\nonumber\\
&&\left.
- {{7051\,{x^8}\,\ln 2}\over {10}} 
+{{3986901\,{x^8}\,\ln 3}\over {1960}} 
+{{696923\,{x^8}\,\ln x}\over {630}} 
\right) \Biggr],
\end{eqnarray}
where $(dJ/dt)_{\rm N}$ is the Newtonian angular momentum
flux expressed in terms of $x$, 
\begin{equation}
\left({dJ_z\over dt}\right)_{\rm N}
={32\over 5}\left(\mu\over M\right)^2 M x^7,
\label{eq:Jnewtonx}
\end{equation}
and the $e$-independent terms in both $\langle dE/dt\rangle$ and 
$\langle dJ/dt\rangle$ are the same and are given by the terms in the
case of circular orbit, Eq.~(\ref{eq:dEdt11}).

\subsection{Slightly eccentric orbit around a Kerr black hole}

Next, we consider a particle 
in slightly eccentric orbit on the equatorial plane 
around a Kerr black hole~\cite{ref:tagoshi}. 
We define the orbital radius $r_0$ and the eccentricity 
in the same way as in the Schwarzschild case by 
\begin{eqnarray}
{\partial(R/r^4)\over \partial r}\bigg|_{r=r_0}=0,
\quad\mbox{and}\quad R(r_0(1+e))=0.
\end{eqnarray}
We also assume $e\ll 1$. 
In this case, $\Omega_\varphi$ is given to $O(e^2)$ by 
\begin{eqnarray}
\Omega_\varphi&=&\Omega_c
\biggl[1 - qv^3
+e^2\left( -{{3}\over {2}} + 
   {9\over 2}v^2 - {9\over 2}qv^3 + 
   3\left( 6 + {q^2} \right)v^4 - 
   60q v^5\right) 
\nonumber\\
&&
+ O(v^6) \biggr]. 
\label{eq:omegaphi}
\end{eqnarray}

We show the energy and angular momentum luminosity
which are accurate to $O(e^2)$ and to $O(x^5)$ beyond
Newtonian order.
\begin{eqnarray}
\left\langle{dE \over dt}\right\rangle
&=&\left({dE\over dt}\right)_{\rm N}
\Bigl\{1-{\frac {1247}{336}}\,{x}^{2}
-{11\over 4}\,q{x}^{3}+4\,{x}^{3}\pi 
-{\frac {44711}{9072}}\,{x}^{4}+{\frac {33}{16}}\,{x}^{4}{q}^{2}
\nonumber\\
&&
-{\frac {59}{16}}\,q{x}^{5}-{\frac {8191}{672}}\,{x}^{5}\pi
+\bigl({\frac {157}{24}}
-{\frac {6781}{168}}\,{x}^{2}
-{\frac {2009}{72}}\,q{x}^{3}
+{\frac {2335}{48}}\,{x}^{3}\pi
\nonumber\\
&&
-{\frac {14929}{189}}\,{x}^{4}
+{\frac {281}{16}}\,{x}^{4}{q}^{2}
-{\frac {2399}{56}}\,q{x}^{5}-{\frac {773}{3}}\,{x}^{5}\pi\bigr){e}^{2}
\Bigr\}, 
\end{eqnarray}

\begin{eqnarray}
\left\langle{dJ_z \over dt}\right\rangle
&=&\left({dJ_z\over dt}\right)_{\rm N}
\Bigl\{
1-{\frac {1247}{336}}\,{x}^{2}
-{11\over 4}\,q{x}^{3}+4\,{x}^{3}\pi 
-{\frac {44711}{9072}}\,{x}^{4}
+{\frac {33}{16}}\,{x}^{4}{q}^{2}
\nonumber\\
&&
-{\frac {59}{16}}\,q{x}^{5}-{\frac {8191}{672}}\,{x}^{5}\pi
+\bigl({\frac {23}{8}}
-{\frac {3259}{168}}\,{x}^{2}
-{\frac {371}{24}}\,q{x}^{3}+{\frac {209}{8}}\,{x}^{3}\pi
\nonumber\\
&&
-{\frac {1041349}{18144}}\,{x}^{4}+{\frac {171}{16}}\,{x}^{4}{q}^{2}
-{\frac {243}{8}}\,q{x}^{5}
-{\frac {785}{6}}\,{x}^{5}\pi \bigr){e}^{2}\Bigr\}.
\end{eqnarray}

\subsection{Circular orbit with small inclination from the 
equatorial plane around a Kerr black hole}

Next, we consider a particle 
in circular orbit with small inclination from the equatorial plane 
around a Kerr black hole~\cite{ref:SSTT}. 
In this case, apart from the energy $E$
and $z$-component of the angular momentum $l_z$, the particle motion 
has another constant of motion; the Carter constant $C$.
The orbital plane of the particle precesses
around the symmetric axis of black hole, and the degree of 
precession is determined by the value of the Carter constant. 
We introduce a dimensionless parameter $y$ defined by 
\begin{eqnarray}
y={C\over Q^2}, \quad\quad Q^2=l_z^2+a^2(1-E^2).
\end{eqnarray}
Given the Carter constant and orbital radius $r_0$, 
the energy and angular momentum is uniquely determined by 
$R(r)=0$, and $\partial R(r)/\partial r=0$.
By solving the geodesic equation with the assumption $y\ll 1$,
we find that $y^{1/2}$ is equal to the inclination angle 
from the equatorial plane. 
The angular frequency $\Omega_\varphi$ is determined 
to $O(y)$ and $O(v^4)$ as
\begin{eqnarray}
\Omega_\varphi=\Omega_c\left[1-qv^3
+{3\over 2}y\left(qv^3-q^2v^4\right)+O(v^6)\right].
\label{eq:inclinedOmega}
\end{eqnarray}

We show the energy and the $z$-component angular flux to $O(v^5)$.
\begin{eqnarray}
\left\langle {dE \over dt} \right\rangle &=&
\left({dE\over dt}\right)_{\rm N}
\biggl( 1 -{1247 \over 336}v^2+4\pi v^3 
 -{73 \over 12}qv^3\left(1-{y\over2}\right)
 -{44711 \over 9072}v^4 \cr
&&+{33 \over 16}q^2v^4-{527 \over 96}q^2v^4 y -{8191 \pi \over 672}v^5
+{3749 \over 336}qv^5\left(1-{y\over2}\right) \biggr). 
\label{eq:Etotincl}
\end{eqnarray} 

\begin{eqnarray}
\left\langle{dJ_z\over dt}\right\rangle
&=&\left({dJ_z\over dt}\right)_{\rm N}
\biggl[
\left(1-{y\over2}\right)-{1247\over336}v^2\left(1-{y\over2}\right)
+4\pi v^3\left(1-{y\over2}\right)
\cr
&&-{61\over12}qv^3\left(1-{y\over2}\right)
-{44711\over9072}v^4\left(1-{y\over2}\right)
+q^2v^4\left({33\over16}-{229\over32}y\right)
\cr
&&-{8191\over672}\pi v^5\left(1-{y\over2}\right)
+qv^5\left({417\over56}-{4301\over224}y\right)
\biggr].
\label{eq:Jtotincl}
\end{eqnarray}

Using Eq.~(\ref{eq:inclinedOmega}), we can express $v$ in terms 
of $x=(M\Omega_\varphi)^{1/3}$ as 
\begin{eqnarray}
v=x\left[1 + {q\over 3}x^3+{1\over 2}y\left(-qx^3 +q^2x^4\right) \right]. 
\end{eqnarray}
We then express Eqs.~(\ref{eq:Etotincl}) and (\ref{eq:Jtotincl}) 
in terms of $x$ as 
\begin{eqnarray}
\left\langle {dE \over dt} \right\rangle &=&
\left({dE\over dt}\right)_{\rm N}
\biggl[1 - \frac{1247}{336}x^2 + \left( 4\,\pi  - \frac{11}{4}q - 
     \frac{47}{24}qy \right) \,x^3 
\nonumber\\
&&
-\frac{44711}{9072}x^4  
+ q^2 x^4\left(\frac{33}{16} - \frac{47}{96}y \right)
-\frac{8191\,\pi }{672}x^5 
\nonumber\\
&&
+qx^5\left(- \frac{59}{16} + 
     \frac{11215}{672}y \right) \biggr],
\\
\left\langle{dJ_z\over dt}\right\rangle
&=&\left({dJ_z\over dt}\right)_{\rm N}
\biggl[
\left(1-{y\over 2}\right) -{1247\over 336}x^2\left(1-{y\over 2}\right)
+4\pi x^3\left(1-{y\over 2}\right)
\nonumber\\
&&-q x^3\left({7\over 2}y+{11\over 4}\left(1-{y\over 2}\right)\right)
-{44711\over 9072}x^4\left(1-{y\over 2}\right)
+q^2x^4\left({33\over 16}-{117\over 32}y\right)
\nonumber\\
&&-{8191\over 672}\pi x^5\left(1-{y\over 2}\right)
+qx^5\left(-{59\over 16}+{687\over 224}y\right)
\biggr].
\end{eqnarray}

\subsection{Absorption of gravitational waves by a black hole}

In this section, we evaluate the energy absorption rate 
by a black hole.
The energy flux formula is given by \cite{ref:TP}
\begin{equation}
\left(dE_{\rm hole}\over dtd\Omega\right)=
\sum_{\ell m}\int d\omega {{_{2}S_{\ell m}^2}\over 2\pi} 
{{128 \omega k (k^2+4 {\tilde \epsilon}^2)(k^2+16 {\tilde \epsilon}^2)
(2 M r_+)^5}\over |C|^2}
|\tilde{Z}^{\rm H}_{\ell m\omega}|^2, 
\label{eq:dedthole}
\end{equation}
where ${\tilde \epsilon}=\kappa/(4r_+)$ and 
\begin{eqnarray}
|C|^2&=&\left((\lambda+2)^2+4 a \omega m -4 a^2 \omega^2\right)
  \left[\lambda^2+36 a \omega m - 36 a^2 \omega^2\right]
\nonumber\\
&&+(2\lambda+3)(96 a^2 \omega^2-48 a \omega m)+144 \omega^2(M^2-a^2).
\end{eqnarray}

In calculating $\tilde{Z}^{\rm H}_{\ell m\omega}$, 
we need to evaluate the upgoing solution
$R^{\rm up}$, and asymptotic amplitude of ingoing, and upgoing 
solution, $B^{\rm inc}$, $B^{\rm trans}$, 
and $C^{\rm trans}$ in Eqs.~(\ref{eq:Rin}) and (\ref{eq:Rup}).
Evaluation of the incident amplitude $B^{\rm trans}$ of 
the ingoing solution is essential in the calculation. 
Poisson and Sasaki~\cite{ref:PoSa} evaluated them, in the 
case of circular orbit around the Schwarzschild black hole, 
up to $O(\epsilon)$ beyond the lowest order, and obtained the 
energy flux at the lowest order, using the method 
we have described in section~\ref{sec:ReggeWheeler}.
Later, Tagoshi, Mano, and Takasugi~\cite{ref:TMT}
evaluated the energy absorption 
rate, in the Kerr case, 
to $O(v^8)$ beyond the lowest order using the method 
in section \ref{sec:MSTformalism}. 
Since the resulting formula is very long and complicated, 
we only show here to $O(v^3)$ beyond the lowest order. 
The energy absorption rate is given by 
\begin{eqnarray}
\left( {dE\over dt}\right)_{\rm H}&&=
{32\over 5}\left(\mu\over M\right)^2 x^{10}\, x^5\, \biggl[
-{1\over 4}\,q-{3\over 4}\,{q}^{3}
+\biggl(-q-{\frac {33}{16}}\,{q}^{3}\biggr){x}^{2} \nonumber\\
&&
+\biggl(2\,q{B_2}+{1\over 2}+{13\over 2}\,
\kappa\,{q}^{2}+{\frac {35}{6}}\,{q}^{2}-{1\over 4}\,{q}^{4}
+{1\over 2}\,\kappa \nonumber\\
&&
+3\,{q}^{4}\kappa+6\,{q}^{3}{B_2}\biggr){x}^{3} \biggr], 
\label{eq:dEdthole}
\end{eqnarray}
where
\begin{equation}
B_n={1\over 2i}\left[\psi^{(0)}\left(3+{n i q\over \sqrt{1-q^2}}\right)
-\psi^{(0)}\left(3-{n i q\over \sqrt{1-q^2}}\right)
\right],
\end{equation}
and $\psi^{(n)}(z)$ is the polygamma function.
We see that absorption effect begins at $O(v^5)$ beyond
the quadrapole formula in the case $q\neq 0$.
If we set $q=0$ in the above formula, 
we have 
\begin{equation}
\left({dE\over dt}\right)_{\rm H}=
\left(dE\over dt\right)_N \left(x^8+O(v^{10})\right), 
\end{equation}
which was obtained by Poisson and Sasaki~\cite{ref:PoSa}.

We note that the leading terms in $(dE/dt)_{\rm H}$
are negative for $q>0$, i.e., the black hole loses the energy if the
particle is co-rotating. This is because of the superradiance for modes
with $k<0$.

%% file: livrev6.texin
\section{Conclusion}
\label{sec:conclusion}

In this article, we described analytical approaches to calculate
gravitational radiation from a particle of mass $\mu$ orbiting a black
hole of mass $M$ with $M\gg\mu$, based upon the perturbation
formalism developed by Teukolsky. A review on the this formalism
was given in section 2.
The Teukolsky equation, which governs the gravitational
perturbation of a black hole, is too complicated to be solved analytically.
Therefore, one has to adopt a certain approximation scheme.
The scheme we employed is the post-Minkowski
expansion, in which all the quantities are expanded in terms of a parameter
$\epsilon=2M\omega$ where $\omega$ is the Fourier frequency of the
gravitational waves. For the source term given by a particle in bound orbit,
this naturally gives the post-Newtonian expansion.

In section 3, we considered the case of a
Schwarzschild background. For a Schwarzschild black hole, one can
transform the Teukolsky equation to the Regge-Wheeler equation.
The advantage of the Regge-Wheeler equation is
that it reduces to the standard Klein-Gordon equation in the flat space
limit, and hence it is easier to understand the post-Minkowskian or
post-Newtonian effects.
Therefore, we adopted this method in the case of a Schwarzschild
background. However, the post-Minkowski expansion of the Regge-Wheeler
equation is not quite systematic, and as one goes to higher orders,
the equations to be solved become increasingly complicated.
Furthermore, for a Kerr background, although one can perform a
transformation similar to the Chandrasekhar transformation,
it can be done only at the expense of losing the reality of the equation.
Thus, the resulting equation is not quite suited 
to analytical treatments.

In section 4, we described a different method, developed by
Mano, Suzuki and Takasugi~\cite{ref:MT,ref:MST}, that directly deals
with the Teukolsky equation and considered the case of a Kerr background
with this method. Although the method is mathematically rather complicated
and it is hard to obtain physical insights into relativistic effects,
it has a great advantage that it allows a systematic post-Minkowski
expansion of the Teukolsky equation, even on the Kerr background.
We gave a thorough review on how this method works and how
it gives a systematic post-Minkowski expansion.

Finally, in section 5, we recapitulated the results of calculations of
the gravitational waves for various orbits that had been obtained
by various authors by the methods described in sections 3 and 4.
These results are useful not only by themselves for the actual case
of a compact star orbiting a supermassive black hole, but also
give us useful insights into higher order post-Newtonian effects
even for a system of equal-mass binaries.

\section{Acknowledgements}

It is our great pleasure to thank Shuhei Mano, Yasushi Mino, 
Takashi Nakamura, Eric Poisson, Masaru Shibata, Eiichi Takasugi
and Takahiro Tanaka, for collaborations and fruitful discussions.
We are grateful to Luc Blanchet and Bala Iyer for useful suggestions
and comments. We are also grateful to Ryuichi Fujita for 
checking formulas in the draft. 
This work was supported in part by Grant-in-Aid for Scientific Research,
Nos.~14047214 and 12640269, of the Ministry of Education, 
Culture, Sports, Science, and Technology of Japan.

%% file: reference.texin
 

%% file: livingrev.bbl
\begin{thebibliography}{199}

\bibitem{ref:ligoweb}
California Institute of Technology, 
``LIGO home page'', (January 2003) 
Cited on 21 January 2003.
[Online HTML document]: http://www.ligo.caltech.edu/ .

\bibitem{ref:virgoweb}
INFN, ``VIRGO home page'', 
(January 2003) Cited on 21 January 2003.
[Online HTML document]: http://www.virgo.infn.it/ .

\bibitem{ref:geoweb}
University of Hanover, 
``GEO 600 home page'', 
(January 2003) Cited on 21 January 2003.
[Online HTML document]: http://www.geo600.uni-hannover.de/ .

\bibitem{ref:tamaweb}
National Astronomy Observatory, Tokyo, 
``TAMA home page'', 
(January 2003) Cited on 21 January 2003.
[Online HTML document]: http://tamago.mtk.nao.ac.jp/ .

\bibitem{ref:uslisaweb}
Jet Propulsion Laboratory, 
``US LISA home page'', 
(January 2003) Cited on 21 January 2003.
[Online HTML document]: http://lisa.jpl.nasa.gov/ .

\bibitem{ref:eurolisaweb}
European Space Agency, 
``European LISA home page'', 
(January 2003) Cited on 21 January 2003.
[Online HTML document]: http://www.estec.esa.nl/spdwww/future/html/lisa.htm .

\bibitem{ref:handbook}
Abramowitz, M., and Stegun, I.A., eds., 
{\it Handbook of Mathematical Functions} (Dover, New York, 1972). 

\bibitem{ref:bar}
Allen, Z.A. et al., 
``First Search for Gravitational Wave Bursts with a Network of Detectors'',
{\it Phys. Rev. Lett.} {\bf 85} 5046 (2000).

\bibitem{ref:tama2}
Ando, M., et al., 
``Stable operation of a 300-m laser interferometer with 
sufficient sensitivity to detect gravitational-wave events
within our galaxy''
{\it Phys. Rev. Lett.} {\bf 86} 3950 (2001). 

\bibitem{ref:AKOP} Apostolatos, T., Kennefick, D., Ori, A., 
and Poisson, E.,
``Gravitational radiation from a particle in circular orbit around a
black hole. III. Stability of circular orbits under radiation reaction'',
{\it Phys. Rev. D} {\bf 47}, 5376 (1993).

\bibitem{ref:BaPr}  Bardeen, J.M., and Press, W.H.,
``Radiation fields in the Schwarzschild background'', 
{\it J. Math. Phys.} {\bf 14}, 7 (1973). 

\bibitem{ref:blanchet} Blanchet, L.,
``Energy losses by gravitational radiation in inspiraling compact
binaries to five halves post-Newtonian order'',
{\it Phys. Rev. D} {\bf 54}, 1417 (1996).

\bibitem{ref:blanreview} Blanchet, L.,
``Gravitational radiation from post-Newtonian sources and
inspiralling compact binaries'',
{\it Living Rev. in Relativity} {\it 5}, (2002), 3. 
[Online Article]: http://www.livingreviews.org/Articles/Volume5/2002-3blanchet/ .

\bibitem{ref:BDIWW} Blanchet, L., Damour, T., Iyer, B.R.,
Will, C.M., and Wiseman, A.G.,
``Gravitational radiation damping of compact binary systems to second
post-Newtonian order'',
{\it Phys. Rev. Lett.} {\bf 74}, 3515 (1995). 

\bibitem{ref:BDI} Blanchet, L., Damour, T., and Iyer, B.R., 
``Gravitational waves from inspiralling compact binaries: Energy loss
and wave form to second post-Newtonian order'',
{\it Phys. Rev. D} {\bf 51}, 5360 (1995). 

\bibitem{ref:BS} Blanchet, L., and Sch\"{a}fer, G.,
``Higher order gravitational radiation losses in binary systems'',
{\it Mon. Not. R. astr. Soc.} {\bf 239}, 845 (1989). 

\bibitem{ref:BS2} Blanchet, L., and Sch\"{a}fer, G., 
``Gravitational wave tails and binary star systems'',
Class. Quantum Grav. {\bf 10}, 2699 (1993). 

\bibitem{ref:breuer} Breuer, R. A.,
{\it Gravitational Perturbation Theory and Synchrotron Radiation:
  Lecture Notes in Physics {\bf 44}}, (Springer-Verlag, Berlin, 1975).

\bibitem{ref:chandratrans} Chandrasekhar, S.,
``On the equations governing the perturbations of the Schwarzschild
black hole'',
{\it Proc. R. Soc. London A } {\bf 343}, 289 (1975). 

\bibitem{ref:chandra} Chandrasekhar, S., 
{\it The Mathematical Theory of Black Holes},
 (Oxford University Press, New York, 1983). 

\bibitem{ref:chrzanowski} Chrzanowski, P.L.,
``Vector potential and metric perturbations of a rotating black hole'',
{\it Phys. Rev. D}, {\bf 11}, 2042 (1975).

\bibitem{ref:three} Cutler, C., Apostolatos, T.A., Bildsten, 
L., Finn, L.S., Flanagan, E.E., Kennefick, D., Markovic, D.M., 
Ori, A., Poisson, E., Sussman, G.J., and Thorne, K.S.,
``The last three minutes: Issues in gravitational wave measurements of
coalescing compact binaries'',
{\it Phys. Rev. Lett.} {\bf 70}, 2984 (1993). 

\bibitem{ref:CFPS} Cutler, C., Poisson, E., Finn, L.S., and
 Sussman, G.J.,
``Gravitational radiation from a particle in circular orbit around a
 black hole. II. Numerical results for the nonrotating case'',
{\it Phys. Rev. D}, {\bf 47}, 1511 (1993).

\bibitem{ref:DaDe1}
Damour, T. and Deruelle, N.,
``Radiation reaction and angular momentum loss 
in small angle gravitational scattering'',
Phys. Lett. {\bf 87A}, 81 (1981). 

\bibitem{ref:DaDe2}
Damour, T. and Deruelle, N.,
``Lagrangien g\'en\'eralis\'e du syst\`eme de deux masses ponctuelles,
\`a l'approximation post-post-newtonienne de la relativit\'e g\'en\'erale'',
{\it C. R. Acad. Sci., Ser. II}, {\bf 293}, 537 (1981).

\bibitem{ref:dixon} Dixon, W.G., 
``Extended bodies in general relativity: Their description and motion'',
in Ehlers, J., ed., 
{\it Isolated Gravitating Systems in General relativity}, 156-219
(North-Holland, Amsterdam, 1979). 

\bibitem{ref:EW} Epstein, R., and Wagoner, R.V., 
``Post-Newtonian generation of gravitational waves'', 
{\it Astrophys J.}, {\bf 197}, 717 (1975).

\bibitem{ref:HTF} Erd\'{e}lyi, A., eds., 
``Higher Transcendental functions Vol. I'', 
(Robert E. Krieger, Florida, 1981).

\bibitem{ref:fackerell} Fackerell, E.D., and Crossman, R.G., 
``Spin-weighted angular spheroidal functions'',
J. Math. Phys. {\bf 9}, 1849 (1977).

\bibitem{ref:Futamase}
Futamase, T.,
``Gravitational Radiation Reaction In The Newtonian Limit'',
{\it Phys.\ Rev.\ D} {\bf 28}, 2373 (1983).

\bibitem{ref:FuSch1}
Futamase, T. and Schutz, B.F.,
``Gravitational Radiation And The Validity Of The Far Zone Quadrupole
 Formula In The Newtonian Limit Of General Relativity'',
{\it Phys. Rev. D} {\bf 28}, 2363 (1983).

\bibitem{ref:FuSch2}
Futamase, T. and Schutz, B.F.,
``Gravitational radiation and the validity of the far-zone quadrupole 
formula in the Newtonian limit of general relativity'',
{\it Phys.\ Rev.\ D} {\bf 32}, 2557 (1985).

\bibitem{ref:galtsov} Gal'tsov, D.V.,
``Radiation reaction in the Kerr gravitational field'',
{\it J. Phys. A.} {\bf 15}, 3737 (1982). 

\bibitem{ref:GMP}  Gal'tsov, D.V.,  Matiukhin, A.A., and Petukhov, V.I., 
``Relativistic corrections to the gravitational radiation of a binary
system and the fine structure of the spectrum'',
{\it Phys. Lett. A} {\bf 77}, 387 (1980). 

\bibitem{ref:gautschi} 
Gautschi, W., 
``Computational aspects of three-term recurrence relations'',
{\it SIAM Rev.} {\bf 9}, 24 (1967). 

\bibitem{ref:spin} Goldberg, J.N., MacFarlane, A.J., 
Newman, E.T., Rohrlich, F., and Sudarshan, E.C.G.,
``Spin-$s$ spherical harmonics and
\rlap{$\partial$}\raise0.3em\hbox{$-$}'',
{it J. Math. Phys.} {\bf 8}, 2155 (1967).

\bibitem{ref:IFA1}
Itoh, Y., Futamase, T., and Asada, H.,
``Equation of motion for relativistic compact binaries with the strong field
 point particle limit I: Formulation, the first post-Newtonian and
multipole terms'',
{\it Phys.\ Rev.\ D} {\bf 62}, 064002 (2000). 

\bibitem{ref:IFA2}
Itoh, Y., Futamase, T., and Asada, H.,
``Equation of motion for relativistic compact binaries with the strong field
 point particle limit : the second and half post-Newtonian order'',
{\it Phys.\ Rev.\ D} {\bf 63}, 064038 (2001).

\bibitem{ref:KenOri} Kennefick, D., and Ori, A., 
``Radiation reaction induced evolution of circular orbits of particles
around kerr black holes'', 
{\it Phys. Rev. D} {\bf 53}, 4319 (1996).

\bibitem{ref:kidd} Kidder, L.E.,
``Coalescing binary systems of compact objects to post-Newtonian 5/2
order. V. Spin effects'',
{\it Phys. Rev. D} {\bf 52} 821 (1995).

\bibitem{ref:KWW} Kidder, L.E., Will, C.M., and Wiseman, A.G.,
``Spin effects in the inspiral of coalescing compact binaries'',
{\it Phys. Rev. D} {\bf 47}, 4183 (1993). 

\bibitem{ref:leaver} Leaver, E.W.,
``Solutions to a generalized spheroidal wave equation: Teukolsky's
equations in general relativity, and the two-center problem in molecular
quantum mechanics'',
{\it J. Math. Phys.} {\bf 27}, 1238 (1986). 

\bibitem{ref:MST} Mano, S., Suzuki, H., and Takasugi, E., 
``Analytic solutions of the Teukolsky equation and their low frequency
expansions'', 
{\it Prog. Theor. Phys.} {\bf 95}, 1079 (1996).

\bibitem{ref:MST2} Mano, S., Suzuki, H., and Takasugi, E., 
``Analytic solutions of the Regge-Wheeler equation and the
post-Minkowskian expansion'',
{\it Prog. Theor. Phys.} {\bf 96}, 549 (1996).

\bibitem{ref:MT} Mano S., and Takasugi, E., 
``Analytic solutions of the Teukolsky equation and their properties'',
{\it Prog. Theor. Phys.} {\bf 97}, 213 (1997). 

\bibitem{ref:supplement}
Mino, Y., Sasaki, M., Shibata, M., Tagoshi, H., and Tanaka, T., 
``Black Hole Perturbation'', 
{\it Prog. Theor. Phys.} Suppl. No. 128, 1 (1997). 

\bibitem{ref:MSTspin} Mino, Y., Shibata, M., and Tanaka, T., 
``Gravitational waves induced by a spinning particle falling into a
rotating black hole'',
{\it Phys. Rev. D} {\bf 53}, 622 (1996).

\bibitem{ref:NOK} Nakamura, T., Oohara, K., and Kojima, Y.,
``General relativistic collapse to black holes and gravitational
waves from black holes'',
{\it Prog. Theor. Phys. Suppl.} {\bf 90}, 1 (1987).

\bibitem{ref:NPS} Narayan, R., Piran, T., and Shemi, A.,
``Neutron star and black hole binaries in the Galaxy'',
{\it Astrophys. J.}, {\bf 379}, L17 (1991).

\bibitem{ref:NP} Newman, E.T., and Penrose, R.,
``An approach to gravitational radiation by a method of
spin-coefficients'', {\it J. Math. Phys.} 
{\bf 7}, 863 (1966).

\bibitem{ref:OTS} Ohashi, A., Tagoshi, H., and Sasaki, M., 
``Post-Newtonian expansion of gravitational waves from a compact star
orbiting a rotating black hole in Brans-Dicke theory: 
Circular orbit case'', 
{\it Prog. Theor. Phys.} {\bf 96}, 713 (1996). 

\bibitem{ref:papa} Papapetrou, A.,
``Spinning test-particles in general relativity'',
{\it Proc. Roy. Soc. London A} {\bf 209}, 243 (1951).

\bibitem{ref:peters} Peters, P.C.,
``Gravitational radiation and the motion of two point masses'',
{\it Phys. Rev.} {\bf 136}, 1224 (1964).

\bibitem{ref:PM} Peters, P.C., and Mathews, J.,
``Gravitational radiation from point masses in a Keplerian orbit'',
{\it Phys. Rev.} {\bf 131}, 435 (1963). 

\bibitem{ref:phinney} Phinney, S.,
``The rate of neutron star binary mergers in the universe 
- Minimal predictions for gravity wave detectors'',
{Astrophys. J.}, {\bf 380}, L17 (1991). 

\bibitem{ref:poisson} Poisson, E.,
``Gravitational radiation from a particle in circular orbit around a
black hole. I. Analytical results for the nonrotating case'',
{\it Phys. Rev. D} {\bf 47}, 1497 (1993).

\bibitem{ref:poisson2} Poisson, E.,
``Gravitational radiation from a particle in circular orbit around a
black hole. IV. Analytical results for the slowly rotating case'',
{\it Phys. Rev. D} {\bf 48}, 1860 (1993). 

\bibitem{ref:poisson3} Poisson, E.,
``Gravitational radiation from a particle in circular orbit around a
black hole. VI. Accuracy of the post-Newtonian expansion'',
{\it Phys.~Rev. D} {\bf 52} 5719 (1995).

\bibitem{ref:PoSa} Poisson, E., and Sasaki, M.,
``Gravitational radiation from a particle in circular orbit around a
black hole. V. Black hole absorption and tail corrections'',
{\it Phys. Rev. D} {\bf 51}, 5753 (1995).

\bibitem{ref:PT} Press, W.H., and Teukolsky, S.A., 
``Perturbations of a rotating black hole II. Dynamical stability of the
Kerr metric'', 
{\it Astrophys. J.} {\bf 185}, 649 (1973). 

\bibitem{ref:RW} Regge, T., and Wheeler, J.A.,
``Stability of a Schwarzschild singularity'',
{\it Phys. Rev.} {\bf 108}, 1063 (1957).

\bibitem{ref:detectorlivrev}
Rowan, S. and Hough, J.,
``Gravitational Wave Detection by Interferometry
(Ground and Space)'' 
{\it Living Reviews in Relativity} {\bf 3} (2000) 3.
[Online Article]: cited on January 2002,
http://www.livingreviews.org/Articles/Volume3/2000-3hough/ .

\bibitem{ref:Ryan} Ryan, F.D., 
``Gravitational waves from the inspiral of a compact object into a
massive, axisymmetric body with arbitrary multipole moments'',
{\it Phys. Rev. D} {\bf 52}, 5707 (1995). 

\bibitem{ref:Ryan2} Ryan, F.D., 
``Effect of gravitational radiation reaction on nonequatorial orbits
around a kerr black hole'',
{\it Phys. Rev. D} {\bf 53}, 3064 (1996).

\bibitem{ref:sasaki} Sasaki, M., 
``Post-Newtonian expansion of the ingoing-wave Regge-Wheeler function'',
{\it Prog. Theor. Phys.} {\bf 92}, 17 (1994). 

\bibitem{ref:SN}
Sasaki, M. and Nakamura, T., 
``A Class Of New Perturbation Equations For The Kerr Geometry'',
{\it Phys. Lett. A} {\bf 89}, 68 (1982).

\bibitem{ref:SN2}
Sasaki, M., and Nakamura, T., 
``Gravitational radiation from a Kerr black hole. I. Formulation and a
method for numerical analysis'',
{\it Prog. Theor. Phys.} {\bf 67}, 1788 (1982). 

\bibitem{ref:shibata} Shibata, M.,
``Gravitational waves by compact stars orbiting around rotating
supermassive black holes'', 
{\it Phys. Rev. D} {\bf 50}, 6297 (1994). 

\bibitem{ref:SSTT} Shibata, M., Sasaki, M., Tagoshi, H.,
 and Tanaka, T.,
``Gravitational waves from a particle orbiting around a rotating black
 hole: Post-Newtonian expansion'',
{\it Phys. Rev. D} {\bf 51}, 1646 (1995).

\bibitem{ref:tagoshi} Tagoshi, H., 
``Post-Newtonian expansion of gravitational waves from a particle in
slightly eccentric orbit around a rotating black hole'',
{\it Prog. Theor. Phys.} {\bf 93}, 307 (1995).

\bibitem{ref:TMT} Tagoshi, H., Mano, S., and Takasugi, E., 
``Post-Newtonian expansion of gravitational waves from a particle in
circular orbits around a rotating black hole: Effects of black hole
absorption'',
{\it Prog. Theor. Phys.} {\bf 98}, 829 (1997). 

\bibitem{ref:TN} Tagoshi, H., and Nakamura, T.,
``Gravitational waves from a point particle in circular orbit around a
black hole: Logarithmic terms in the post-Newtonian expansion'',
{\it Phys. Rev. D} {\bf 49}, 4016 (1994). 

\bibitem{ref:TS} Tagoshi, H., and Sasaki, M., 
``Post-Newtonian expansion of gravitational waves from a particle in
circular orbit around a Schwarzschild black hole'',
{\it Prog. Theor. Phys.} {\bf 92}, 745 (1994).

\bibitem{ref:TSTS} Tagoshi, H., Shibata, M., Tanaka, T.,
and Sasaki, M.,
``Post-Newtonian expansion of gravitational waves from a particle in
circular orbits around a rotating black hole: 
Up to $O(v^8)$ beyond the quadrupole formula'',
{\it Phy. Rev. D} {\bf 54}, 1439 (1996).

\bibitem{ref:tamafirst}
Tagoshi, H., et al.,
``First search for gravitational waves from inspiraling compact binaries
using TAMA300 data'',
{\it Phys. Rev. D} {\bf 63} 062001 (2001). 

\bibitem{ref:TMSS} Tanaka, T., Mino, Y., Sasaki, M.,
and Shibata, M., 
``Gravitational waves from a spinning particle in circular orbits around
a rotating black hole'',
{\it Phys. Rev. D} {\bf 54}, 3762 (1996). 

\bibitem{ref:TTS} Tanaka, T., Tagoshi, H., and Sasaki, M.,
``Gravitational waves by a particle in circular orbits around a
Schwarzschild black hole: 5.5 Post-Newtonian formula'',
{\it Prog. Theor. Phys.} {\bf 96}, 1087 (1996). 

\bibitem{ref:Teu}  Teukolsky, S.A.,
``Perturbations of a rotating black hole I. Fundamental equations for
gravitational, electromagnetic, and neutrino-field perturbations'',
{\it Astrophys. J.} {\bf 185}, 635 (1973).

\bibitem{ref:TP} Teukolsky, S.A., and Press, W.H.,
``Perturbations of a rotating black hole III. Interaction of the hole
with gravitational and electromagnetic radiation'',
{\it Astrophys. J.} {\bf 193}, 443 (1974).

\bibitem{ref:membran} Thorne, K.S., Price, R.M., and MacDonald, D., 
{\it Black Holes: The Membrance Paradigm} (Yale University Press,
New Haven, 1986). 

\bibitem{ref:WaWi} Wagoner, R.V., and Will, C.M.,
``Post-Newtonian gravitational radiation from orbiting
point masses'',
{\it Astrophys. J.} {\bf 210}, 764 (1976). 

\bibitem{ref:Wald} Wald, R.M.,
``Construction of Solutions of Gravitational, Electromagnetic, 
or Other Perturbation Equations from Solutions of Decoupled Equations'',
{\it Phys. Rev. Lett.} {\bf 41}, 203 (1978).

\bibitem{ref:wald} Wald, R.M.,
``Gravitational spin interaction'',
{\it Phys. Rev. D} {\bf 6}, 406 (1972).

\bibitem{ref:WW} Will, C.M., and Wiseman, A.G.,
``Gravitational radiation from compact binary systems: Gravitational
wave forms and energy loss to second post-Newtonian order'',
{\it Phys. Rev. D} {\bf 54}, 4813 (1996). 

\bibitem{ref:wisem} Wiseman, A.G.,
``Coalescing binary systems of compact objects to
(post)$^{5/2}$-Newtonian order. IV. The gravitational
wave tail'',
{\it Phys. Rev. D} {\bf 48}, 4757 (1993).

\bibitem{ref:Zerilli} Zerilli, F.J., 
``Gravitational field of a particle falling in a Schwarzschild 
geometry analyzed in tensor harmonics'', 
{\it Phys. Rev.} {\bf D2}, 2141 (1970). 

\end{thebibliography}
